\documentclass[aps,prc,amssymb,twocolumn,nofootinbib,10pt,preprintnumbers]{revtex4-1}
\pdfoutput=1

\usepackage[utf8]{inputenc}
\usepackage{graphicx}
\usepackage{xcolor}
\usepackage{soul}
\usepackage{dcolumn}
\usepackage{dsfont}
\usepackage{relsize}
\usepackage{scalerel}
\usepackage{bm}
\usepackage{array}
\usepackage{amssymb}
\usepackage{bbold}
\usepackage[]{complexity}
\usepackage[version=3]{mhchem}
\usepackage{slashed}
\usepackage[bookmarks=false]{hyperref}

\newcommand{\pvec}[1]{\vec{#1}\mkern2mu\vphantom{#1}}

\begin{document}

\preprint{ MITP/21-008}

\title{
Beam-normal single-spin asymmetry in
\\
elastic scattering of electrons from a spin-0 nucleus}

\author{Oleksandr Koshchii$^{a}$}
\author{Mikhail Gorchtein$^{b}$}
\author{Xavier Roca-Maza$^{c}$}
\author{Hubert Spiesberger$^{a}$}

\affiliation{$^{a}$PRISMA$^+$Cluster of Excellence,
    Institut f\"ur Physik,
	Johannes Gutenberg-Universit\"at, D-55099 Mainz, Germany}
\affiliation{$^{b}$PRISMA$^+$Cluster of Excellence,
    Institut f\"ur Kernphysik,
	Johannes Gutenberg-Universit\"at, D-55099 Mainz, Germany}
\affiliation{$^{c}$ Dipartimento di Fisica, Universit\'a degli Studi
    di Milano, Via Celoria 16, I-20133 Milano, Italy \\
    and INFN, Sezione di Milano, Via Celoria 16, I-20133 Milano,
    Italy}

\date{\today}

\begin{abstract}
We study the beam-normal single-spin asymmetry (BNSSA) in
high-energy elastic electron scattering from several spin-0
nuclei. Existing theoretical approaches work in the plane-wave formalism and predict the BNSSA to 
scale as  $\sim A/Z$ with the atomic number $Z$ and nuclear mass number $A$.
While this prediction holds for light and intermediate nuclei, a striking disagreement in both the sign and the magnitude of BNSSA
was observed by the PREX collaboration for $^{208}$Pb, coined the ``PREX puzzle".
To shed light on this disagreement, we go beyond the plane-wave approach which neglects 
Coulomb distortions known to be significant for heavy nuclei. 
We explicitly investigate the dependence of BNSSA on $A$ and $Z$ by 
i) including  inelastic intermediate states' contributions 
into the Coulomb problem in the form of an optical
potential, ii) by accounting for the experimental information on the
$A$-dependence of the Compton slope parameter, and iii) giving a thorough account of 
the uncertainties of the calculation. Despite of these improvements, 
the PREX puzzle remains unexplained. We discuss further strategies to resolve this riddle. 
\end{abstract}

\maketitle

\section{Introduction}
\label{Sec:Intro}

Corrections due to two-photon exchange (TPE) Feynman diagrams
for electron scattering have received considerable interest
in recent years
\cite{CarlsonTPE2007,ArringtonTPE2011,GorchteinTPE2014,
AfanasevTPE2017,KoshchiiAsymmetry2017,Pulak.ChiralTPE.2020,
ZhouTPE2020,RachekTPE2015,RimalTPE2017,HendersonTPEOlympus2017,
QattanTPE2018,QattanTPE2020}, primarily in the context of the
discrepancy of the experimental results
\cite{JonesFF2000,GayouFF2002}
for the electric-to-magnetic form factor ratio, which sometimes
is referred to as the proton form factor puzzle. There are strong indications 
\cite{AhmedTPE2020} that this puzzle can be
resolved by a proper inclusion of TPE in the experimental
analysis. In view of the interest in TPE, beam- and target-normal
single spin asymmetries (SSAs) in elastic electron-nucleus
($e N$) scattering regained attention of theorists
\cite{AfanasevSSA2004,AfanasevBNSSA2004,GorsteinBNSSA2006,
PasquiniSSA2004,GorsteinBNSSA2004,CooperCoulDist2005,
BorisyukBNSSA2006,GorchteinDispersive2007,GorsteinBNSSA2008,
CarlsonSSA2017,KoshchiiTNSSAen2018,KoshchiiBnSSAmup2019}.
It has been known for several decades that these transverse
asymmetries are T-odd observables which, in the absence of CP
violation, are sensitive to the imaginary part of the scattering
amplitude \cite{RujulaSSA1971}. The T-even one-photon exchange
amplitude (in the plane-wave Born approximation, PWBA) is purely real, 
and it is the T-odd imaginary (absorptive) part of the TPE amplitude that gives rise to
nonzero transverse asymmetries. 

The measurement of the BNSSA 
(Mott asymmetry, Sherman function and analyzing power are alternative names which are more common for low-energy electron scattering)
has been part of the parity-violation program over the past
two decades
\cite{SampleBNSSA2001,A4Mass2005,G0BNSSA2007,A4Capozza2007,
G0BNSSA2011,AbrahamyanBnSSA2012,GouBnSSA2020,Androic:2020rkw,
EsserBNSSA2018,Esser:2020vjb}. Parity violation is
observed in $eN$ scattering when the incoming electron beam
is longitudinally polarized. Measurements of the respective
parity-violating (PV) asymmetry have far-reaching applications,
including precision tests of the Standard Model
\cite{AndroicQweak2013,AndroicQWeak2018,BeckerP22018} and
studies of the nuclear structure \cite{HorowitzPREX2012,
AbrahamyanPREX2012,YangEWprobes2019,Koshchii:2020qkr}.
Typical values of the PV asymmetry range from parts per million
to parts per billion, several orders of magnitude below BNSSA,
hence a thorough control of this source of a potentially
significant systematical uncertainty associated with an
unknown transverse component of the electron beam polarization
has become a must-do in the analyses of PV electron scattering.
Thanks to the fact that these experiments are designed for
measuring the much smaller PV asymmetry, in the past decades
good-quality data of the BNSSA have become available
in a variety of kinematic regions and for a variety of targets.

The general theoretical treatment of transverse asymmetries
in high-energy ($E_{\rm b}\gtrsim 1$~GeV) elastic $e N$
scattering is a highly challenging task. The two
approaches that have been pursued in the literature in
this energy range are i) solving the Dirac equation for the
electron moving in the Coulomb field of an infinitely heavy
nucleus in the distorted-wave Born approximation (DWBA) upon neglecting nuclear and hadronic excitations of
the intermediate states \cite{CooperCoulDist2005};
and ii) including the latter only in the approximation of
the two-photon exchange \cite{GorsteinBNSSA2008}, disregarding
multi-photon exchange effects. 
The former approach enables one to accurately account for Coulomb
distortion effects which scale with the nuclear charge,
$Z \alpha$, and thus are important for electron scattering
from heavy nuclei. While this mechanism
dominates at electron energies in the few-MeV range,
its contribution to BNSSA drops with the electron energy, and for GeV electrons the
inelastic hadronic contribution exceeds the former by several
orders of magnitude \cite{GorsteinBNSSA2008}.
When compared with the corresponding scattering data from
Jefferson Lab \cite{AbrahamyanBnSSA2012} and
MAMI \cite{EsserBNSSA2018,Esser:2020vjb},
the second approach has been quite successful for light and
intermediate-mass nuclei, e.g. ${}^{4}$He, ${}^{12}$C, ${}^{28}$Si,
and ${}^{90}$Zr, while a stark disagreement between theory and
experiment for BNSSA on the ${}^{208}$Pb target
\cite{AbrahamyanBnSSA2012}, sometimes called the ``PREX puzzle",
was observed. This disagreement indicates that the theoretical
calculation of Ref.~\cite{GorsteinBNSSA2008} may miss some
important nuclear contributions which become important for very
heavy nuclei, while only playing a minor role otherwise.
One such effect might be the exchange of many soft Coulomb photons
on top of the two-photon exchange which may lead to a substantial
modification of the leading-order result.

In this article, we join the two aforementioned approaches. 
We include the contribution
of the inelastic hadronic states as an optical potential entering
the Dirac equation and study the interplay of the Coulomb distortion
and two-photon exchange within one formalism. We also improve the
existing calculations by using a more extensive database for
experimental information on Compton scattering on nuclei. We
use this information to extract the dependence of the optical
potential on the nuclear mass number.

\section{Dirac Coulomb problem at relativistic energies}
\label{Sec:partial_waves}

We consider elastic scattering of an electron of mass $m$ by a
spin-$0$ nucleus of mass $M$,
\begin{equation}
\label{Eq:sc_process}
    e^-(k_1, S_i) + N(p_1) \rightarrow e^-(k_2, S_f) + N(p_2) \, ,
\end{equation}
where $k_1$ ($k_2$) and $p_1$ ($p_2$) denote the four-momenta of
the initial (final) electron and initial (final) nucleus, and
$S_i$ ($S_f$) describes the spin projection of the initial (final)
electron along the considered axis.

The beam-normal single-spin asymmetry is defined as
\begin{equation}
\label{Eq:bnssa}
\begin{split}
    B_n & \equiv \frac{\sigma_{\uparrow} - \sigma_{\downarrow}}
    {\sigma_{\uparrow} + \sigma_{\downarrow}},
\end{split}
\end{equation}
where $\sigma_{\uparrow}$ ($\sigma_{\downarrow}$) represents
the $eN$ scattering cross section for electrons with spin parallel
(anti-parallel) to the normal vector $\xi^\mu$ given by
\begin{equation}
\label{Eq:pol_vector}
    \xi^\mu = (0, \vec{\xi}), \ \ \ \ \vec{\xi} \equiv
    \frac{\vec{k}_1 \times \vec{k}_2}{|\vec{k}_1 \times \vec{k}_2|}.
\end{equation}

In order to account for Coulomb distortion and inelastic
intermediate excitations in the considered scattering process,
we solve the relativistic Dirac equation\footnote{We use natural
units throughout this paper.}:
\begin{equation}
\label{Eq:dirac_eq}
\begin{split}
\Big(- i \vec{\alpha}\cdot \vec{\nabla} + \beta m + V_{\rm c}
+ i \beta V_{\rm abs} \Big) \Psi(\vec{r}\,)
= E\Psi(\vec{r}\,) \, ,
\end{split}
\end{equation}
where $\vec{\alpha} = \gamma_0 \vec{\gamma}$ and $\beta = \gamma_0$
are Dirac matrices, and $E$ the electron energy in the center of
mass reference frame related (neglecting the electron mass) to
the laboratory energy $E_{\rm b}$ by $E  =E_{\rm b} /
\sqrt{1 + 2E_{\rm b}/M}$.

The Coulomb potential $V_{\rm c}({r})$ corresponds to the
nuclear charge distribution which is known from electron
scattering experiments \cite{Vries1987}. The absorptive
potential $V_{\rm abs}({r}, E)$ represents the contribution
of the inelastic hadronic excitations in the two-photon exchange
diagram, as discussed in detail in Sec.~\ref{Sec:AbsPot}. 
The inclusion of the absorptive component of the potential
in the Dirac problem is the main novel feature of this work.
Note that the form of this potential, $ i \beta V_{\rm abs}$, is specific to the problem at hand: 
an absorptive  potential of the form $ i \tilde V_{\rm abs}$ only contributes to $B_n$ at 
higher order in $\alpha$, exceeding the precision goal of this study. 
{
Spherically symmetric $V_{\rm c}({r})$ and
$V_{\rm abs}({r}, E)$ should be expected for spin-0 nuclei, and we use this assumption throughout this paper.}

For electron scattering in a central field, the solution of
the Dirac equation can be expanded in spherical waves
\cite{SalvatELSEPA2005},
\begin{align}
\label{Eq:dirac_sol}
\Psi_{\kappa, m_z}(\vec{r}\,) = \frac{1}{r}
\left(
  \begin{array}{cc}
    P_{\kappa}(r) \, \Omega_{\kappa, m_z}(\theta, \phi)
    \\
    i Q_{\kappa}(r) \, \Omega_{- \kappa, m_z}(\theta, \phi)
  \end{array}
\right) \, ,
\end{align}
where $\Omega_{\kappa, m_z}(\theta, \phi)$ are 2-component spherical
spinors. The relativistic quantum number $\kappa$ takes values
$\kappa_1$ and $\kappa_2$ given by
\begin{align}
\label{Eq:kappa}
\begin{cases}
        \kappa_1 = - (j + 1/2) \ \ \mathrm{if} \ \ j = l + 1/2 \, ,
        \\
        \kappa_2 = + (j + 1/2) \ \ \mathrm{if} \ \ j = l - 1/2 \, ,
\end{cases}
\end{align}
where $l, j$, and $m_z$ are the orbital angular momentum,
total angular momentum, and total angular momentum projection
quantum numbers, respectively.

The radial functions $P_{\kappa}(r)$ and $Q_{\kappa}(r)$
satisfy the following coupled system of differential equations:
\begin{align}
\label{Eq:radial_comp}
  \frac{dP_{\kappa}}{dr}
  & =
  - \frac{\kappa}{r} P_{\kappa}
  + \Big( E - V_{\rm c} + i V_{\rm abs} + m \Big)
  Q_{\kappa} \, , \nonumber
  \\
  \frac{dQ_{\kappa}}{dr}
  & =
  - \Big( E - V_{\rm c} - i V_{\rm abs} - m \Big)
  P_{\kappa}
  + \frac{\kappa}{r} Q_{\kappa}
  \, .
\end{align}

We normalize the spherical waves such that the radial function
$P_{\kappa}(r)$ oscillates asymptotically with unit amplitude,
\begin{align}
\label{Eq:upper_radial_func}
  P_\kappa(r \rightarrow \infty)
  = \sin \left( kr - l \frac{\pi}{2}
  - \eta \ln 2kr + \delta_\kappa \right),
\end{align}
where $k$ is the electron's wave number and $\eta=-Z \alpha E/k$
is the relativistic Sommerfeld parameter. The scattering phase
shift $\delta_\kappa$ is obtained by requiring continuity of
the radial function $P_\kappa(r)$ and its derivative at large
distance $r_{\rm m}$ (matching distance), at which the numerical
solution of Eq.~(\ref{Eq:radial_comp}) is matched to the known
analytical solution of the Dirac equation for a point-like
Coulomb potential, $V_{\rm pc} (r)\!=\!- Z \alpha/r$.
{
The matching at large distances is justified
by the fact that both the absorptive potential and the short
range part of the Coulomb potential can be neglected beyond
$r_{\rm m}$. As a result, the solution of the point-like
Coulomb potential provides the proper asymptotic behavior.}

The absorptive potential, while having a shorter range than the
Coulomb one, turns out to extend to distances of the order of
the inverse electron mass $1/m\sim400$~fm, and the respective
computation becomes cumbersome (details are discussed in
Sec.~\ref{Sec:AbsPotResult}). To perform the numerical
calculation, we use the ELSEPA package \cite{SalvatELSEPA2005,
SalvatRadial2019}, properly modified to include the absorptive
potential.

Knowledge of the phase shift enables one to determine the direct
and spin-flip scattering amplitudes, $f(\theta)$ and $g(\theta)$,
respectively, in terms of which the beam-normal SSA is
given by
\begin{equation}
\label{Eq:sherman}
\begin{split}
B_n
=
i \frac{f(\theta) g^* (\theta) - f^{*} (\theta) g(\theta)}
{|f(\theta)|^2 + |g (\theta)|^2}
=
\frac{2\,{\rm{Im}} \left[f^{*} (\theta) g(\theta)\right]}
{|f(\theta)|^2 + |g (\theta)|^2}
\, .
\end{split}
\end{equation}
These amplitudes admit the following partial-wave
expansions:
\begin{align}
\label{Eq:scat_amp}
f (\theta)
& = \frac{1}{2ik} \sum \limits_{l=0}^{\infty}
\big[ (l+1) \left( e^{2 i \delta_{\kappa_1}} - 1 \right)
\nonumber \\
& \qquad \qquad \quad
+ l \left( e^{2 i \delta_{\kappa_2}} - 1 \right) \big]
P_l (\cos \theta) \, ,
\nonumber \\
g (\theta)
& = \frac{1}{2ik} \sum \limits_{l=0}^{\infty}
\left[ e^{2 i \delta_{\kappa_2}} - e^{2 i \delta_{\kappa_1}}
\right] P_l^1 (\cos \theta) \, ,
\end{align}
where $P_l (\cos \theta)$ and $P_l^1 (\cos \theta)$ are Legendre
and associated Legendre polynomials.  The series in
Eqs.~(\ref{Eq:scat_amp}) is singular at $\theta = 0$ leading
to a slow convergence when approaching that limit. The
convergence of the series can be accelerated by using the
reduced series method suggested by Yennie et al.\ in
Ref.~\cite{Yennie.PhaseShiftSum.1954}. This method prescribes
to reduce the degree of the singularity of the original series
by expanding $(1-\cos \theta)^n f(\theta)$ and
$(1-\cos \theta)^n g(\theta)$ into analogous sums over Legendre
polynomials. The new sums converge more quickly, however the
extraction of the original amplitudes requires to divide by a
factor $(1 - \cos \theta)^n$. As a result, for forward scattering
the use of too many reductions becomes unstable. We found the
optimal number of reductions to be $n\!=\!2$.

\section{Absorptive potential from the two-photon exchange}
\label{Sec:AbsPot}

\subsection{Elastic $eN$ scattering}
\label{Sec:general_amp}

We turn to a field-theoretical description of the $eN$ scattering
process to deduce the explicit form of the potentials in
Eq.~(\ref{Eq:dirac_eq}). In the absence of P- and CP-violation,
the invariant amplitude describing the scattering process
Eq.~(\ref{Eq:sc_process}) for a spin-0 nucleus has two terms,
\cite{GorsteinBNSSA2008},
\begin{align}
\label{Eq:general_ampl}
T
=\frac{e^2}{|t|}
\bar{u}(k_2)
\left[ m A_1 (s, t) +
(\slashed{p}_1 + \slashed{p}_2)A_2(s, t) \right]
u(k_1)
\, ,
\end{align}
with the usual Mandelstam invariants $t = (k_1 - k_2)^2$ and
$s = (k_1 + p_1)^2$, and two scalar amplitudes $A_1$ and $A_2$.
The initial and final electron Dirac spinors are denoted by
$u(k_1)$ and $u(k_2)$, respectively.

In the static approximation, $|t|\ll s,\,M^2,\,E^2$, relativistic
electron-nucleus scattering reduces to the problem of potential
scattering of a relativistic electron in the field of a static
nucleus. In the static limit, $p_1 = p_2 = (M,0)$, we can
rewrite Eq.~(\ref{Eq:general_ampl}) as
\begin{equation}
\label{Eq:general_ampl_b}
\begin{split}
T
= 2M
u^\dagger(k_2)
\left[ \frac{e^2}{|t|} \left(\frac{m \beta}{2 M}
A_{1} + A_{2}\right)
\right] {u}(k_1)
\,.
\end{split}
\end{equation}

\begin{figure}[t]
  \includegraphics[width=0.9\linewidth]{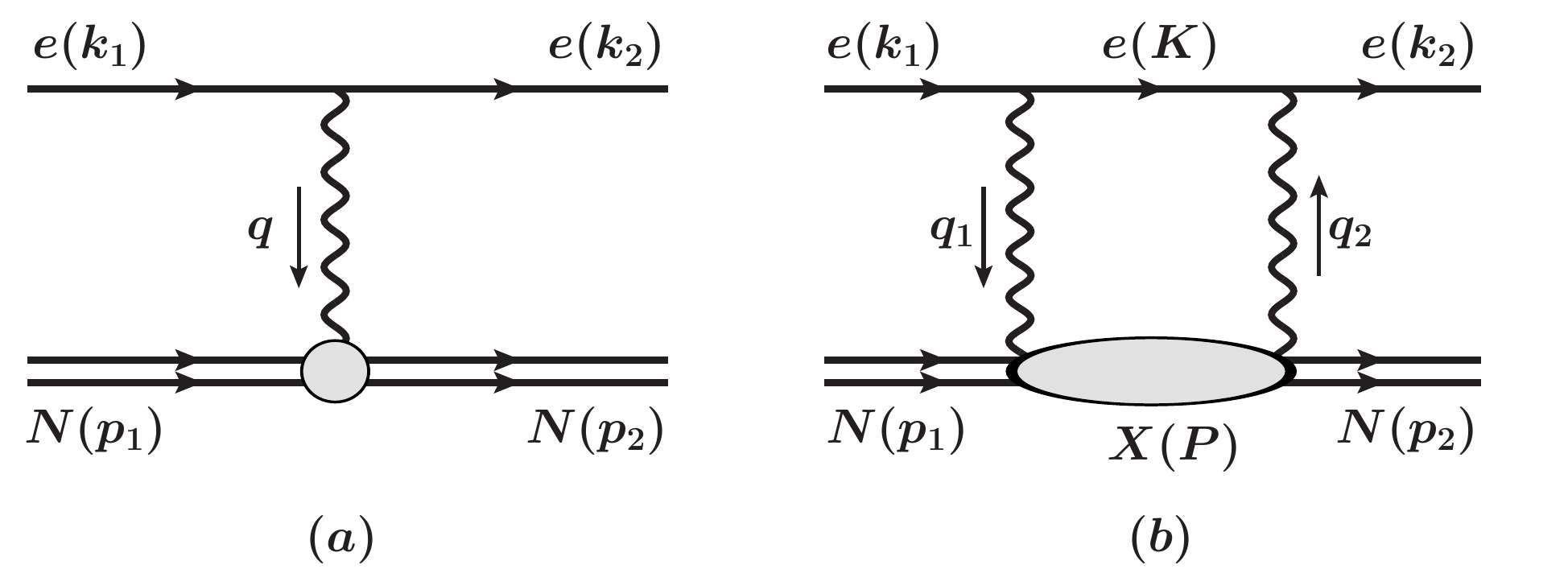}
  \caption
  {
  (a) One- and (b) two-photon exchange diagrams for elastic
  electron-nucleus scattering.
  }
  \label{fig:tpe}
\end{figure}

To leading order in $Z\alpha$, the electron-nucleon interaction
proceeds via the exchange of a virtual photon,
cf.\ Fig.~\ref{fig:tpe}(a), and only the amplitude $A_2$
survives at this order,
\begin{align}
\label{Eq:amplitudes_ope}
A_{1}^{1 \gamma} = 0 \, , \quad
A_{2}^{1 \gamma} = Z F_{\rm ch} (t) \, .
\end{align}
Here, $F_{\rm ch}$ denotes the nuclear charge form factor,
which is related to the spatial distribution of the nuclear charge
$\rho_{\rm ch}(r)$ by a three-dimensional Fourier transform,
\begin{equation}
\label{Eq:fch}
F_{\rm ch}(t) =
\int\!\rho_{\rm ch}(r) e^{-i\vec{q}\cdot\vec{ r}} d^{3}\vec{r},
\hspace{5pt} {\rm with} \hspace{3pt}
|\vec{q}\,|\!\equiv\!\sqrt{|t|} \, ,
\end{equation}
with the normalization $\!\int\!\rho_{\rm ch}(r) d^{3}\vec{r}\!=\!1$.

The T-odd observable $B_n$ is determined by the imaginary
part of the interference of $A_1$ and $A_2$ \cite{RujulaSSA1971}.
An imaginary part, $\mathrm{Im}A_{1}$, for elastic $eN$ scattering,
i.e.\ for $s>M^2$, $t<0$, appears first at next-to-leading order
in $Z\alpha$ in the two-photon exchange contribution depicted in
Fig.~\ref{fig:tpe}(b).

The absorptive and Coulomb components of the total potential which
enter the Dirac equation \eqref{Eq:dirac_eq}, $V_{\rm abs}$ and
$V_{\rm c}$, are related to $\mathrm{Im}A_{1}$ and $ A_{2}$
through a three-dimensional Fourier transform of the first and
second terms in square brackets of Eq.~(\ref{Eq:general_ampl_b}).
Including the leading nonvanishing terms in the perturbative
expansion of these amplitudes we obtain
\begin{align}
&V_{\rm c}({r})
=
- \int \frac{d^3 \vec{q} }{(2 \pi)^3}\,
\frac{e^2}{\vec{q}^{\,2}} A_{2}^{1 \gamma} \,
e^{i \vec{ q} \cdot \vec{ r}}
=
- Z \alpha
\int d^3 \pvec{r}' \,
\frac{\rho_{\rm ch}(\pvec{r}')}{|\vec{ r} - \pvec{r}'|},
\label{Eq:coul_pot}
\\
&V_{\rm abs} ({r}, E_{\rm b})
=
- \int \frac{d^3 \vec{q}}{(2 \pi)^3} \,
\frac{e^2}{\vec{q}^{\,2}} \frac{m}{2 M}
\mathrm{Im}A_{1}^{2 \gamma} \, e^{i \vec{ q} \cdot \vec{ r}}.
\label{Eq:abs_potential}
\end{align}
Other contributions, e.g.\ recoil corrections and higher-order
contributions to $A_2$, are neglected here. In the next section
we are going to study the perturbative result for the two-photon
exchange diagram in order to determine an explicit ansatz for
the absorptive potential.

\subsection{Imaginary part of the two-photon exchange amplitude}
\label{Sec:abs_tpe}

The imaginary part of the two-photon exchange amplitude, displayed
in Fig.~\ref{fig:tpe}(b), is given by
\begin{align}
\label{eq:absorptive_tpe}
    \mathrm{Im}T_{2 \gamma}
    = e^4 \int & \frac{d^3 \vec{K}}{(2 \pi)^3 2 E_K}
    \frac{2\pi L_{\alpha \beta} W^{\alpha \beta}}{Q^2_1 Q^2_2},
\end{align}
where the momenta are defined as shown in Fig.~\ref{fig:tpe},
with $Q_{1,2}^2\!=\!-q_{1,2}^2\!=\!-(k_{1,2}\!-\!K)^2$,
and $E_K$ and $\vec{K}$ the energy and three-momentum of
the intermediate electron inside the loop, respectively.
The leptonic tensor $L_{\alpha\beta}$ reads
\begin{align}
\label{eq:lept_tensor}
    & L_{\alpha \beta} = \bar{u}(k_2) \gamma_\alpha
    (\slashed{K} + m) \gamma_\beta u(k_1) \, ,
\end{align}
and the doubly virtual Compton scattering (VVCS) tensor
$W^{\alpha \beta}$ is defined as {
\begin{align}
\label{eq:hadr_tensor_a}
    W^{\alpha \beta}
    & \equiv \frac{1}{4\pi} \sum \limits_X \left< p_2| J^{\dag}_\alpha (0)|X(P) \right>
     \left< X(P)| J_{\beta} (0)|p_1 \right>  \\
    & \hspace{1.4cm} \times (2 \pi)^4 \delta^4 \big( p_2 + q_2 - P \big) \nonumber \\
    & = \frac{1}{4\pi}
    \int d^4x e^{iq_2 x}
    \left< p_2| [J^{\alpha \dag} (x), J^{\beta} (0)] |p_1 \right> \, ,
\label{eq:hadr_tensor_b}
\end{align}
where $\sum_X$ in Eq.~(\ref{eq:hadr_tensor_a}) includes the phase-space integral 
$\int d^3\vec{P}/((2\pi)^3 2E_P)$. We note that in order to
get from Eq.~(\ref{eq:hadr_tensor_a}) to Eq.~(\ref{eq:hadr_tensor_b})
one can apply a translation to the current operator,
$J_\alpha (x) = e^{i P x} J_\alpha(0) e^{-i P x}$, and
$(2 \pi)^4 \delta^4 ( p_2 + q_2 - P) =
\int d^4x e^{i(p_2 + q_2 - P)x}$.
The matrix element of the hadronic current for the elastic
intermediate state is given by
$\left<P| J_\beta (0)|p_1 \right> = (P + p_1)_\beta \,
Z F_{\rm ch} (Q_1^2)$.}

To compute the imaginary part of the TPE diagram and perform a
systematic study of its uncertainties, we note that the result
of the contraction with the leptonic tensor can be decomposed
into two parts,
\begin{align}
\label{Eq:tensor_contr}
    L_{\alpha \beta} W^{\alpha \beta}
    = m \, \bar{u}(k_2) u(k_1) \, {\cal{A}}_1
    + \bar{u}(k_2) \big(\slashed{p}_1 + \slashed{p}_2\big)
    u(k_1) \, {\cal{A}}_2 \, ,
\end{align}
where ${\cal{A}}_1$ and ${\cal{A}}_2$ are analytical functions of
$t$, $Q_1^2$, $Q_2^2$, $W^2=(p_1+q_1)^2$ and $s$. With this notation, a
straightforward connection to the amplitude $A_1$ can be made,
\begin{align}
\label{Eq:ImA12gamma}
    {\rm Im}A_1^{2\gamma}
    &=
    \frac{\alpha|t|}{2 \pi}\int
    \frac{\vec K^2d|\vec K|d\Omega_K}{E_KQ_1^2Q_2^2}
    {\cal{A}}_1(t,Q_1^2,Q_2^2,W^2,s) \, .
\end{align}
In the following we will obtain the long-range (i.e.\ low-$t$)
behavior of ${\rm Im}A_1^{2\gamma}$, adequate for devising the form of the absorptive potential via Eq. (\ref{Eq:abs_potential}).

To that end, we follow
Refs.~\cite{AfanasevBNSSA2004,AfanasevSSA2004,GorsteinBNSSA2006}
which observed that $B_n$ is logarithmically enhanced in the
kinematical regime $m^2 \ll |t| \ll s$ due to the collinear photon singularity.
The integrals over the solid angle that are prone to this enhancement read
\begin{align}
&I_0 =
\frac{|t|\vec K^2}{2\pi}\int\frac{d\Omega_K}{Q_1^2Q_2^2}
\approx
\ln\frac{|t|}{m^2} \, ,
\nonumber\\
&I_1 =
\frac{E|\vec K|}{\pi}\int\frac{d\Omega_K}{Q_1^2}
= \frac{E|\vec K|}{\pi}\int\frac{d\Omega_K}{Q_2^2}
\approx \ln\frac{4E^2}{m^2} \, ,
\label{eq:MasterIntegrals}
\end{align}
with the energies $E = (s-M^2)/(2\sqrt{s})$ and $E_K =
(s-W^2)/(2\sqrt{s})$ defined in the center-of-mass frame,
and $|\vec K| = \sqrt{E_K^2 - m^2}$. Here we have listed only
the leading behavior in the limit where $\sqrt{|t|}$ and $E$
are large compared with the electron mass $m$. The exact
expressions are given in the Appendix~\ref{App:master_int}.
For the values we are interested in, $|t| \approx 0.01$~GeV$^2$,
the first logarithm is of order 10. The second logarithm is of
order 25 for $E$ in the GeV range, but is suppressed by an
explicit factor $|t|$ with respect to the former.
This hierarchy defines our approximation scheme:
\begin{align}
{\rm Im}A_1^{2\gamma}
&=
\alpha \int\frac{d|\vec K|}{E_K}
\Big[ {\cal{A}}_1^{(0)}(t) I_0
+ |t|\frac{E_K}{2E}{\cal{A}}_1^{(1)}(t)I_1
+ \dots\Big] \, ,
\label{eq:ImA1:approx}
\end{align}
where terms denoted by dots are doubly suppressed: they
contain one power of $t$ and no large logarithm. To
arrive at this result we have used an expansion in small
photon virtualities,
\begin{align}
{\cal{A}}_1(t, Q_1^2, Q_2^2, W^2, s)
= {\cal{A}}_1^{(0)}(t)
+ \frac{Q_1^2+Q^2_2}{2}{\cal{A}}^{(1)}_{1}(t)
+ \dots \, ,
\end{align}
where we show explicitly only the dependence of ${\cal{A}}_1^{(0)}$
and ${\cal{A}}_1^{(1)}$ on $t$, while their dependence on
the other four invariant variables is implicitly assumed. For
consistency, we will only keep the ``strong'' $t$-dependence
in ${\cal{A}}_1^{(0)}$, ${\cal{A}}_1^{(1)}$, e.g.\ an
exponential or the nuclear charge form factor, but will
neglect power corrections $\sim t/M^2$, $t/s$, $t/E^2$. In the
literature only ${\cal{A}}_1^{(0)}(t)$ has been obtained in the
near-forward limit. In this work we include the second term
and use it to estimate the uncertainty induced by the
approximations used.

Next we proceed to derive explicit expressions for
${\cal{A}}_1^{(0)}$ and ${\cal{A}}_1^{(1)}$. The optical
theorem relates them to the total cross sections for virtual
photoabsorption at the first step. The $t$-dependence is
reconstructed at the second step from the measured differential
cross section for real Compton scattering. This two-step procedure
requires that we operate with the Compton amplitudes which are
well-defined in both
\\
a) the forward scattering limit, described by $t=0$ and finite
$Q_1^2 = Q_2^2 \equiv Q^2$, and
\\
b) the real Compton scattering limit, described by $Q_1^2 =
Q_2^2 = 0$ and finite $t$.
\\
The general virtual Compton tensor $W^{\alpha\beta}$ for a
spinless target consists of five independent Lorentz structures,
$\tau_i^{\alpha \beta}$, ($i = 1, \ldots, 5$)
\cite{Tarrach1975,DrechselVVCS-spin0-1997,Lensky-VVCS-Spin0-2018}
multiplied by respective scalar amplitudes
$\mathcal{F}_i (t, Q_1^2, Q_2^2, W^2)$.
In the approximation scheme we work in, the number of structures
that contribute is further reduced upon neglecting terms that
vanish in both the forward and the real Compton scattering
limits. This restricts our consideration to just two structures
$\tau_i^{\alpha \beta}$ (in the original enumeration of
Ref.~\cite{Tarrach1975}):
\begin{align}
\label{Eq:RCS}
     W^{\alpha \beta}
     = &
     \tau_1^{\alpha \beta}{\rm Im}\, \mathcal{F}_1
     + \tau_3^{\alpha \beta}{\rm Im}\, \mathcal{F}_3 \, ,
     \nonumber \\
     \tau_1^{\alpha \beta}
     = &
     (q_1 \cdot q_2) g^{\alpha \beta}
     - q_1^{\alpha} q_2^{ \beta} \, ,
     \nonumber \\
     \tau_3^{\alpha \beta}
     = &
     (\bar{p} \cdot \bar{q})^2 g^{\alpha \beta}
     - (\bar{p} \cdot \bar{q})
       (\bar{p}^\beta q_1^\alpha + \bar{p}^\alpha q_2^{\beta})
     \nonumber \\
     & + (q_1 \cdot q_2) \bar{p}^\alpha \bar{p}^\beta
     \, ,
\end{align}
with $\bar{p} = (p_1 + p_2)/2$ and $\bar{q} = (q_1 + q_2)/2$.
Other structures (explicitly provided in Ref.~\cite{Tarrach1975})
surviving in the forward limit for virtual photons can always
be expressed as linear combinations of $\tau_{1,3}$.

By contracting the leptonic and virtual Compton tensors,
we find an explicit expression for ${\cal{A}}_1$,
\begin{align}
\label{Eq:h0h1}
     {\cal{A}}_1 (t)
     & = {\rm Im}\,\mathcal{F}_1
     \left[ (Q_1^2 + Q_2^2) (\lambda - 2) -  t \right]
     \hspace{1.9cm} \\
     & + {\rm Im}\,\mathcal{F}_3 \nonumber  \big[
     \left(4 (\bar{p} \cdot \bar{q}) (\bar{p} \cdot \bar{k})
     - \bar{p}^2 (q_1 \cdot q_2) \right)  \\
     & \hspace{1.5cm}
     \times (\lambda - 1)
     - 2 (\bar{p} \cdot \bar{q})^2 (\lambda - 2) \big]
     \, ,
     \nonumber
\end{align}
where $\bar{k} = (k_1 + k_2)/2$ and
\begin{align}
    \lambda =
    \frac{2(s-M^2)(s-W^2)-(Q_1^2+Q_2^2)(s+M^2)}{2(s-M^2)^2}
    \, .
    \nonumber
\end{align}

The forward limit of the amplitudes $\mathcal{F}_{1,3}$ is
determined by the usual structure functions $F_1$ and $F_2$,
\begin{align}
\label{Eq:amplitudes}
     Q^2 {\rm Im}\,\mathcal{F}_1(0,Q^2,Q^2,W^2)
     & = F_1 -\frac{1}{2x_{\rm Bj}}F_2 \equiv F_L \, ,
     \nonumber \\
     Q^2 {\rm Im}\,\mathcal{F}_3(0,Q^2,Q^2,W^2)
     & = - \frac{1}{(\bar{p}\cdot \bar{q})} F_2 \, ,
\end{align}
where $x_{\rm Bj} = Q^2 / (2 \bar{p} \cdot \bar{q})$. The
structure functions are related to the transverse and
longitudinal inelastic virtual photoabsorption cross sections
$\sigma_T$ and $\sigma_L$ via
\begin{align}
\label{Eq:struc_f}
    F_1(W^2, Q^2)
    & =
    \frac{W^2 - M^2}{8 \pi^2 \alpha} \sigma_{T} (W^2, Q^2) \, ,
    \\
    F_2(W^2, Q^2)
    & = \frac{W^2 - M^2}{8 \pi^2 \alpha}
    \frac{Q^2 \, (\bar{p}\cdot \bar{q})}%
         {(\bar{p}\cdot \bar{q})^2+Q^2 M^2}
    \nonumber \\
    & \times
    \Big(\sigma_{T} (W^2, Q^2) + \sigma_{L} (W^2, Q^2) \Big) \, .
    \nonumber
\end{align}
Expanding $\sigma_{T,L}$
at $Q^2 = 0$ we obtain the following expressions for
${\cal{A}}_1^{(0)}$ and ${\cal{A}}_1^{(1)}$:
\begin{align}
\label{Eq:H0}
    {\cal{A}}_1^{(0)}(0)
    &=
    \frac{M}{2 \pi^2 \alpha} E_K \frac{\omega}{E}
    \sigma_T^0(\omega) \, ,
    \\
    {\cal{A}}_1^{(1)}(0)
    &=
    \frac{M}{2 \pi^2 \alpha} \bigg[
    \left( -\frac{2}{\omega} + \frac{3}{2 E_{\rm b}}
    - \omega \frac{(E_{\rm b}+M)}{2M E_{\rm b}^2} \right)
    \sigma_T^0 (\omega)
    \nonumber
    \\
\label{Eq:H1}
    & \hspace{1.1cm}
    + E_K\frac{\omega}{E} \sigma'_{T}(\omega)
    + 2 \omega \sigma'_{L}(\omega)
    \bigg] \, ,
\end{align}
where $E_{\rm b} = (s-M^2)/(2M)$ and instead of the variable
$W^2$ we used $\omega = (W^2-M^2)/(2M)$. Moreover,
$\sigma_T^0(\omega) \equiv \sigma_T(\omega, Q^2 = 0)$
and $\sigma'_{T,L}(\omega) \equiv d\sigma_{T,L}/dQ^2(\omega,Q^2=0)$.

\renewcommand{\arraystretch}{1}
\begin{table*}[t!]
\centering
\caption{
\label{tab:slope}
Average values and corresponding theoretical uncertainties for
the Compton slope parameter $B$ extracted from Compton scattering
data of Refs.~\cite{Aleksanian:1986hb} and \cite{Criegee:1977uf}.
For nuclei with $A\ge12$, $d\sigma/dt(0)=a+\sigma_{\rm inc}$ is fixed to the value
reported in Ref.~\cite{Criegee:1977uf}.
{
The 5~GeV results are used for all nuclei except for $^{4}$He.}
}
\begin{tabular}{m{0.07\textwidth}|m{0.14\textwidth}|m{0.14\textwidth}|m{0.14\textwidth}|
                m{0.14\textwidth}|m{0.14\textwidth}|m{0.14\textwidth}}
\hline\hline
\rule[-2mm]{0mm}{6mm}
       & \multicolumn{3}{c|}{$\omega \simeq 3$\,GeV}
       & \multicolumn{3}{c}{$\omega = 5$\,GeV}
       \\
\hline
\rule[-2mm]{0mm}{6mm}
Target
&  $a$ [$\mu$b/GeV$^{2}$]
&  $\sigma_{\rm inc}$ [$\mu$b/GeV$^{2}$]
&  $B$ [GeV$^{-2}$]
&  $a$ [$\mu$b/GeV$^{2}$]
&  $\sigma_{\rm inc}$ [$\mu$b/GeV$^{2}$]
&  $B$ [GeV$^{-2}$]
\\
\hline
$^{4}$He   & $13.1\pm2.5$ & $0.0$       & $10.0\pm3.6$  & ---              & ---           & --- \\
$^{12}$C   & $111.9$      & $0.0$       & $7.2\pm2.5$   & $89.7\mp0.8$     & $2.9\pm0.8$   & $10.0\pm2.1$\\
$^{27}$Al  & $523.0$      & $0.0$       & $12.1\pm2.1$  & $402.8\mp1.1$    & $2.2\pm1.1$   & $8.1\pm1.9$\\
$^{49}$Ti  & ---          & ---         & ---           & $1210.9\mp1.5$   & $9.1\pm1.5$   & $18.6\pm2.1$\\
$^{64}$Cu  & $2664.0$     & $0.0$       & $16.5\pm13.8$ & $2022.4\mp1.9$   & $9.6\pm1.9$   & $14.6\pm2.3$\\
$^{109}$Ag & $8406.0$     & $0.0$       & $26.4\pm3.5$  & $6096.1\mp2.8$   & $24.9\pm2.8$  & $26.0\pm2.7$\\
$^{197}$Au & ---          & ---         & ---           & $20589.2\mp17.4$ & $40.8\pm17.4$ & $56.7\pm8.2$ \\[0.3ex] \hline\hline
\end{tabular}
\end{table*}

The $t$-dependence of the Compton amplitudes can be retrieved
from experimental studies of the differential cross section
for Compton scattering. Measurements are available at high
energies, $E \sim 3-5$~GeV and at low $-t$, $0.001 < -t <
0.06$~GeV$^{2}$, see Refs.~\cite{Aleksanian:1986hb,
Criegee:1977uf}. In this kinematic range,
\begin{align}
    \frac{d\sigma}{dt}
    &\approx
    \frac{\pi \alpha^2 M^2\omega^2}{16}
    \left|{\rm Im}\mathcal{F}_3\right|^2\left(1+R^2\right) \, ,
\label{eq:ComptonCS}
\end{align}
where the terms suppressed with powers of $t$ were neglected,
and we defined $R = \left|{\rm Re} \mathcal{F}_3\right| /
\left|{\rm Im}\mathcal{F}_3\right|$.
{
The data
follow an exponential fall-off,
\begin{align}
\label{Eq:Comp_data_fit}
    \frac{d \sigma}{dt}(\omega,t)
    & = 
    a e^{- B |t|} F^2_{\rm ch}(t) +  \sigma_{\rm inc}
    \, .
\end{align}
$F_{\rm ch}(t)$ is the
nuclear charge form factor. Depending on the nucleus, we adopt
a two-parameter Fermi model ($^{197}$Au,
$^{109}$Ag, and $^{64}$Cu),
a Fourier-Bessel model ($^{49}$Ti, $^{27}$Al, and $^{12}$C),
or a sum of Gaussians ($^{4}$He) \cite{Roca_Maza_2017,Vries1974,Vries1987}.
A (small) incoherent contribution $\sigma_{\rm inc}$ was added to improve the description of the data
around the first diffraction minimum and above.
In the $t$-range of interest this
contribution is a slowly-varying function of $t$ which can be approximated by a polynomial.
In practice, we found that only the constant term is reliably constrained by the data. This is related
to the rather small range of $t$ where data are available, as well as large uncertainties
at the largest values of the momentum transfer.
We perform the fit using the $3$~GeV and $5$~GeV data of
Ref.~\cite{Criegee:1977uf} with $B$ and $\sigma_{\rm inc}$ as free parameters, and
fix the normalization of the coherent contribution $a$ such that the sum $a+\sigma_{\rm inc}$
reproduces the values of $d\sigma/dt(0)$ reported in Ref.~\cite{Criegee:1977uf}.

We find the incoherent contributions to be irrelevant for the description of the
data at 3~GeV and set $\sigma_{\rm inc}$ to 0 (cf.\ second column of Table~\ref{tab:slope}).
For the 5~GeV data, instead, its inclusion greatly improves the overall fit due to a larger
measured $t$-range.
Importantly, however, whether including or excluding the incoherent contribution from the fit
barely affects the extracted value of $B$, as the latter is determined by low-$t$ data.
For the $^4$He data of Ref.~\cite{Aleksanian:1986hb}, we treat $a$ as a free parameter.
The extracted values for $a$, $B$,
and $\sigma_{\rm inc}$ are listed in Table~\ref{tab:slope}.
We display the fit of the
$5$~GeV data of Ref.~\cite{Criegee:1977uf}
in Fig.~\ref{fig:Bfit-vs-data}}.

In the literature, BNSSA measurements have been reported for the following
spin-0 nuclei: $^{4}$He, $^{12}$C, $^{28}$Si, $^{40}$Ca,
$^{48}$Ca, $^{90}$Zr and $^{208}$Pb. To obtain the slope
parameter $B$ for $^{28}$Si, $^{40}$Ca, $^{48}$Ca, $^{90}$Zr and
$^{208}$Pb for which no direct data is available, we use the
values obtained for nuclei with the closest atomic weight in
Table~\ref{tab:slope}. {
More specifically, we
use the values of $B$ from $^{27}$Al for $^{28}$Si, $^{49}$Ti for $^{40,48}$Ca, $^{109}$Ag for $^{90}$Zr, and $^{197}$Au for $^{208}$Pb. 
Values of $B$ obtained from the fit to Compton data at $\omega = 3$~GeV and $\omega = 5$~GeV are compatible with each other (where a comparison is possible). We use the more precise values from the $\omega = 5$~GeV fit for all nuclei except for $^{4}$He where only data at $\omega = 3.3$~GeV are available.

\begin{figure}[b]
  \includegraphics[width=1\linewidth]{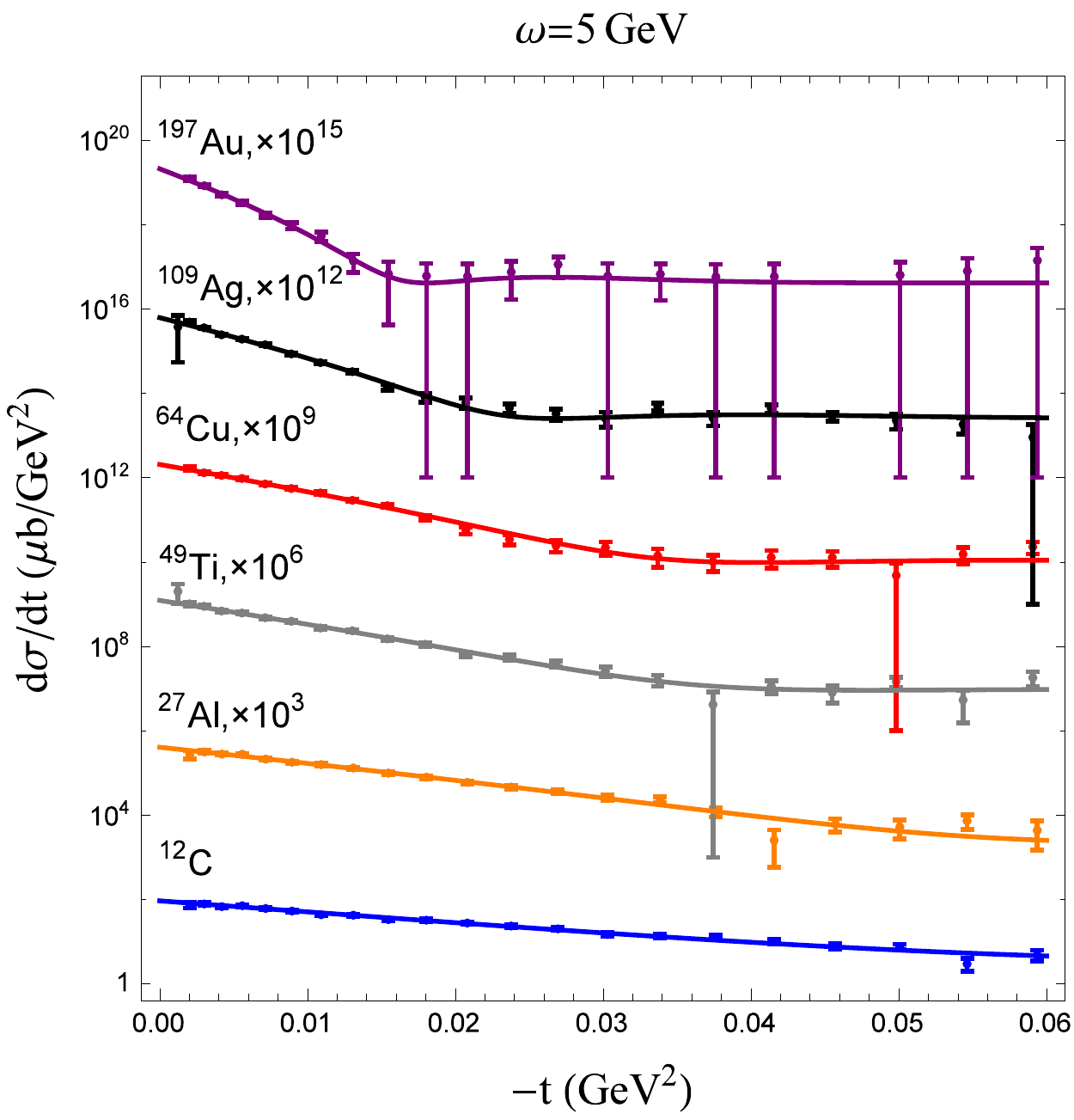}
  \caption{
  Compton scattering cross section data \cite{Criegee:1977uf}
  used to determine the Compton slope $B$ compared with our
  fit.
  }
  \label{fig:Bfit-vs-data}
\end{figure}

We finally reconstruct the $t$-dependence of the imaginary
part of the Compton amplitude using Eqs.~(\ref{eq:ComptonCS},
\ref{Eq:Comp_data_fit}),
\begin{align}
\label{Eq:ImF3tdep}
    {\rm Im}{\cal{F}}_3(\omega,t)
    =
    {\rm Im}{\cal{F}}_3(\omega,0) \,
    e^{- \frac{B |t|}{2}}
    F_{\rm ch}(t)\sqrt\frac{1+R^2(\omega,0)}{1+R^2(\omega,t)} \, ,
\end{align}
and estimate the ratio $R(\omega,t)$ in a Regge model. For
the latter, we use a recent Regge fit \cite{Gorchtein:2011xx}
of the total photoabsorption cross section measured for several
nuclei \cite{Caldwell:1973bu,Caldwell:1978ik,Bianchi-Photoabsorption-1996}.
The total cross section was fitted by a sum of a Pomeron
and a Reggeon exchange,
\begin{align}
    \sigma^{tot}_{\gamma A}(\omega)
    =
    c_P^A(\omega/\omega_0)^{\alpha_P(0)-1}
    + c_R^A(\omega/\omega_0)^{\alpha_R(0)-1} \, ,
\end{align}
with $\omega_0 = 1$~GeV and linear Regge trajectories $\alpha_i(t) = \alpha_{0i} +
\alpha'_it$. The intercepts are $\alpha_{0P} = 1.097$,
$\alpha_{0R} = 0.5$ and the slopes $\alpha'_{P} = 0.25$~GeV$^{-2}$,
$\alpha'_{R} = 0.9$~GeV$^{-2}$ \cite{Gorchtein:2011xx}.
The values of $c_{P,R}^A$ for carbon, aluminum, copper and lead
are listed in Table I of Ref.~\cite{Gorchtein:2011xx}. From the
optical theorem, $\sigma_{\gamma A}^{tot}(\omega) \propto
{\rm Im}{\cal{F}}^{\gamma A\to\gamma A}_3$. For a given Regge
exchange contribution to an amplitude, its real and imaginary
parts follow from the known phase of the Regge propagator,
\begin{align}
  {{P}}_R
  \sim
  e^{i\pi\alpha_R(t)} + \xi,\quad\xi = \pm 1 \, .
\end{align}
For Compton scattering only the exchange with natural parity,
$\xi = +1$, contributes to the spin-independent channel,
and $R_i^2(t) = \cot^2[\pi\alpha_i(t)/2]$ with $i=P,R$.
We found that removing the effect of the real part of the
amplitude from the differential cross section to obtain the
$t$-dependence of the imaginary part, Eq.~(\ref{Eq:ImF3tdep}),
is equivalent to a change in the slope parameter $B$ of
$\approx2.2$~GeV$^{-2}$. We include this effect as an
uncertainty of $B$, in addition to those of the fit listed
in Table~\ref{tab:slope}, and use a simple exponential times
charge form factor ansatz for the $t$-dependence of the amplitude
${\rm Im}{\cal{F}}_3$.

Finally, our ansatz for the $t$-dependence of the coefficients
${\cal{A}}_1^{(0)}(t)$ and ${\cal{A}}_1^{(1)}(t)$ in front
of the large logarithms at small momentum transfer $t$ reads
\begin{align}
\label{Eq:ImF3ansatz}
    {\cal{A}}_1^{(0,1)}(t)
    &=
    {\cal{A}}_1^{(0,1)}(0) \, e^{- \frac{B |t|}{2}}
    F_{\rm ch}(t) \, .
\end{align}
While the ansatz for ${\cal{A}}_1^{(1)}(t)$ is motivated by the
continuity of the function ${\cal{A}}(t,Q_1^2,Q_2^2,W^2,s)$
in $t$, $Q^2$ near the forward limit and near the real photon
point, the quality of this approximation is hard to estimate.
We therefore assign a 100\% uncertainty to the
${\cal{A}}_1^{(1)}$ contribution.

\subsection{Absorptive potential}
\label{Sec:AbsPotResult}

The results of Eqs.~(\ref{eq:ImA1:approx}, \ref{Eq:H0},
\ref{Eq:H1}), and (\ref{Eq:ImF3ansatz}) can be used to compute
the absorptive potential given by Eq.~(\ref{Eq:abs_potential}).
{
Using the hierarchy introduced
in Eq.~(\ref{eq:ImA1:approx}), we split the absorptive
potential into two parts,
$V_{\rm abs} ({r}, E_{\rm b}) = V_{\rm abs}^{(0)} ({r}, E_{\rm b})
+ V_{\rm abs}^{(1)} ({r}, E_{\rm b})$, and find the leading
and subleading contributions to be given by
\begin{align}
\label{Eq:abs_potential_final}
    &V_{\rm abs}^{(0)}
     = c_0 \int\limits_{\omega_\pi}^{E_{\rm b}} d\omega \omega \,
    \sigma_{T}^{0} (\omega) \int\limits_0^\infty
    dq \, j_0(q r) F_{\rm ch}(q^2) e^{- \frac{B}{2}q^2} I_0 \, , \nonumber \\
    &V_{\rm abs}^{(1)}
     =  \frac{c_0}{2}\int\limits_0^\infty
    dq \, q^2 j_0(q r) F_{\rm ch}(q^2) e^{- \frac{B}{2}q^2}  \int\limits_{\omega_\pi}^{E_{\rm b}} d\omega I_1\nonumber\\
   &\times
    \bigg\{ \left[\frac{3}{2 E_{\rm b}}- \frac{2}{\omega}
    - \frac{ \omega (E_{\rm b}+M)}{2M E_{\rm b}^2} \right]
    \sigma_{T}^0
    + \frac{\omega(E_{\rm b}-\omega)}{E_{\rm b}}
    \sigma'_{T} \bigg\} \, ,
\end{align}
with $c_0\!=\!- \alpha m / (2 \pi^3 E_{\rm b})$,
$q \equiv |\vec{q}\,|$, $j_0(qr)$ the Bessel function
of order zero, and $\omega_{\pi} = m_{\pi} + m_{\pi}^2/(2M)$
the laboratory frame photon energy at the pion photoproduction threshold.
The expressions for $I_0$ and $I_1$ are provided in the
Appendix~\ref{App:master_int}.

We note that the $\omega$-weighting $\sim\omega\sigma_T^0(\omega)$ in $V_{\rm abs}^{(0)}$, together
with the overall $1/E$-weighting in $c_0$, puts the emphasis on the photoabsorption in the hadronic energy range. Nuclear
photoabsorption occurs at much lower energies and its contribution to the leading term is suppressed.

In this article, we focus on the evaluation of contributions to
the absorptive potential coming from photoabsorption in the hadronic region. 
In the nucleon resonance region and slightly above, the total nuclear
photoabsorption cross section is assumed to approximately scale
with the atomic weight $A$ as $\sigma_{T}(\omega)
\approx A \sigma_{T, \gamma p}(\omega)$, where
$\sigma_{T, \gamma p}(\omega)$ is the real photoabsorption
cross section of the proton. For the
evaluation of $\sigma_{T,\gamma p}^0$ and
$\sigma'_{T, \gamma p}$, we use the parametrization of Ref.~\cite{ChristyFit2010}. We point out that
$\sigma'_{L, \gamma p}(\omega)$ is zero in this parametrization,
hence this contribution was omitted in
Eq.~(\ref{Eq:abs_potential_final}).

The naive linear $A$-scaling disregards the shadowing at higher energies and
anti-shadowing in the resonance region (cf.\ Fig.~10 of
Ref.~\cite{Bianchi-Photoabsorption-1996}).
Nevertheless, since Eq.~(\ref{Eq:abs_potential_final}) operates with the integrated cross section
rather than the cross section itself, the two effects should largely cancel out justifying our approximation.
A comprehensive study of specifically nuclear effects in photoabsorption, from the giant resonance to shadowing
and anti-shadowing at hadronic energies, is postponed to a future work.
}

\begin{figure}[t]
  \includegraphics[width=1\linewidth]{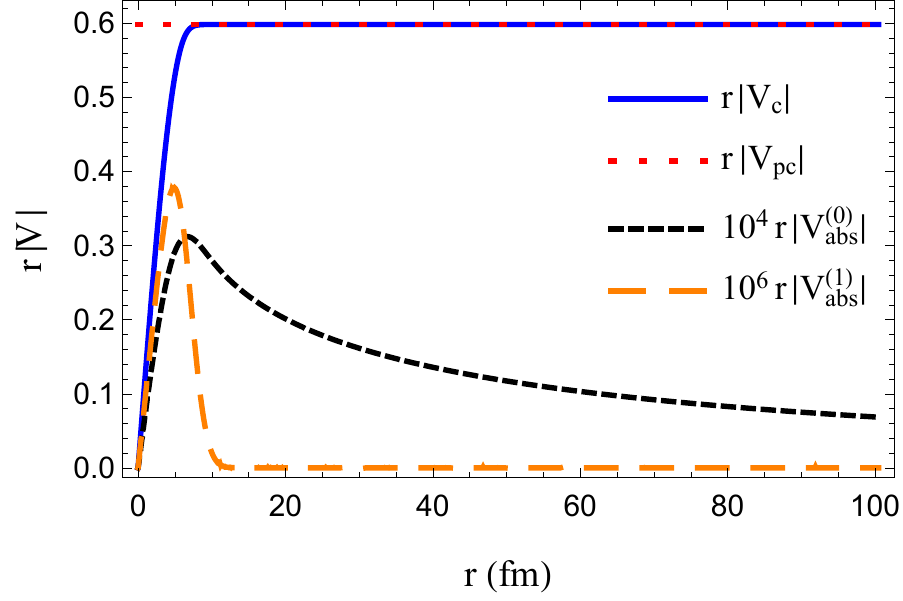}
  \caption{
  The $r$-dependence of weighted potentials $r |V(r)|$ for
  $^{208}$Pb as a function of $r$ in units of Fermi.
  The point-charge Coulomb potential is shown by a red
  dotted horizontal line at $r|V_{\rm pc}(r)| = Z \alpha$.
  The Coulomb potential of the empirical charge distribution
  corresponds to the blue solid curve. The black and orange
  dashed curves show the leading and subleading contributions
  to the absorptive potential, correspondingly, for
  $E_{\rm b} = 1.063$~GeV.
  }
  \label{fig:abs_pot}
\end{figure}

In Fig.~\ref{fig:abs_pot}, we display the result of a
numerical evaluation of the two-fold integrals in
Eq.~(\ref{Eq:abs_potential_final}) which determine the
leading (black dashed curve) and subleading (orange dashed
curve) contributions to the absorptive potential for
$^{208}$Pb at $E_{\rm b} = 1.063$~GeV. We compare the result
with the Coulomb potential (blue solid curve) of the lead
nuclear charge distribution of Ref.~\cite{Vries1987} and with
the Coulomb potential for a point-like charge (red dotted curve).
The potentials for the point-like and the empirical charge
distributions approach each other just outside the r.m.s.~radius,
which is $\sim 5.5$~fm for lead. We observe that the absorptive
potential is rather small (note the scale factor $10^4$ in front
of the leading contribution to $V_{\rm abs}$)
and has a finite range ($r V_{\rm abs}(r\to\infty)\to0$).
However, it extends far outside the nuclear charge distribution
as it is sensitive to scales up to the electron's Compton wave
length $\sim 1/m \sim 400$~fm. This property results in a
large matching distance needed for a precise evaluation of
$B_n$. The matching distance $r_{\rm m}$ is the distance at
which the total interaction potential
$V(r) = V_{\rm ch}(r) \pm i V_{\rm abs}(r)$ has reached
(within a given precision) its asymptotic value
$V(r_{\rm m}) = V_{\rm{pc}}(r_{\rm m})$,
beyond which $V_{\rm abs}$ can be set equal to zero. The
determination of the matching distance is of crucial
importance for our calculation since $r_{\rm m}$ is the
distance where the numerical solution of the Dirac equation
is matched to the known analytical solution with
$V(r) = V_{\rm pc}(r)$. We observe that the CPU time for
the numerical calculation grows approximately linearly
with $r_{\rm m}$, and a proper balance between precision
and computing time had to be found.

We studied the dependence of predictions for $B_n$ on the
matching distance for electron scattering from $^{208}$Pb
at $E_{\rm beam} = 1.063$~GeV. In previous calculations for
the Coulomb problem with a nuclear charge distribution of a
typical radius $\lesssim6$~fm and without including an
absorptive potential, a matching distance of $r_{\rm m}
\sim 15$~fm had been used \cite{CooperCoulDist2005}. However,
we found that the relative uncertainty of our calculation for
$B_n$ at $\theta = 5^\circ$ can not be expected to be better than
$10^{-3}$ if $r_{\rm m}$ is chosen smaller than $120$~fm.
In addition, the precision of the calculation becomes worse as
the scattering angle increases. The results of our calculation
for $B_n$, which are presented in Sec.~\ref{Sec:results}, are
obtained with $r_{\rm m} = 606$~fm. Such a matching distance
represents a compromise between achieving the necessary
numerical precision and keeping the calculation time under
control. With $r_{\rm m} = 606$~fm, the relative intrinsic
numerical uncertainty of our prediction for $B_n$ is well
below $\sim 1\%$ in the range of momentum transfers considered
in this paper, independently of the target and beam energy.

\begin{figure}[t]
  \includegraphics[width=1\linewidth]{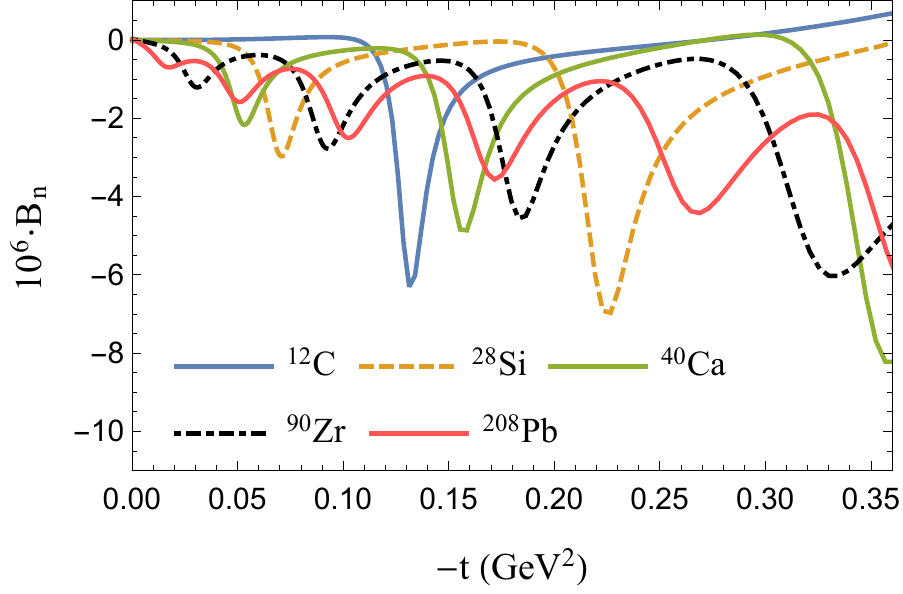}
  \caption{
  BNSSA for elastic electron scattering from
  $^{12}$C (solid blue curve), $^{28}$Si (dashed orange curve),
  $^{40}$Ca (solid green curve), $^{90}$Zr (dashed-dotted black
  curve), and $^{208}$Pb (solid red curve) versus momentum transfer
  squared $|t|$ at $E_{\rm b} = 953$~MeV for the case when only
  elastic intermediate-state contributions are taken into account.
  For the distribution of the nuclear charge, $\rho_{\rm ch}(r)$,
  we use an experimental fit of the world data on elastic
  electron-nucleus scattering parametrized in the form of a sum
  of Gaussians ($^{12}$C, $^{28}$Si, $^{40}$Ca, $^{208}$Pb) or
  Fourier-Bessel ($^{90}$Zr) as reported in Ref.~\cite{Vries1987}.}
  \label{fig:SSA_no_vabs}
\end{figure}

\begin{figure*}[t]
  \includegraphics[scale=0.9]{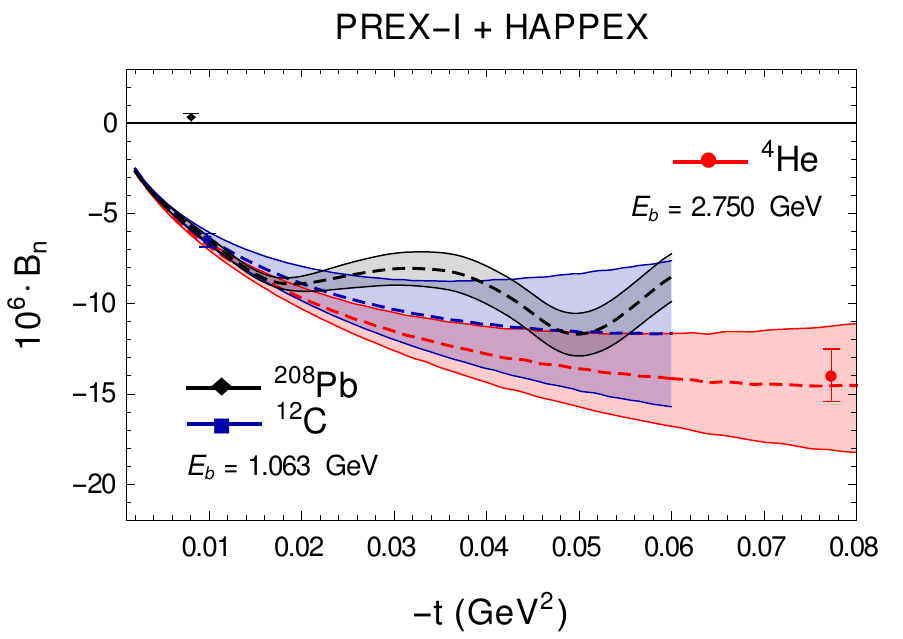}
  \includegraphics[scale=0.9]{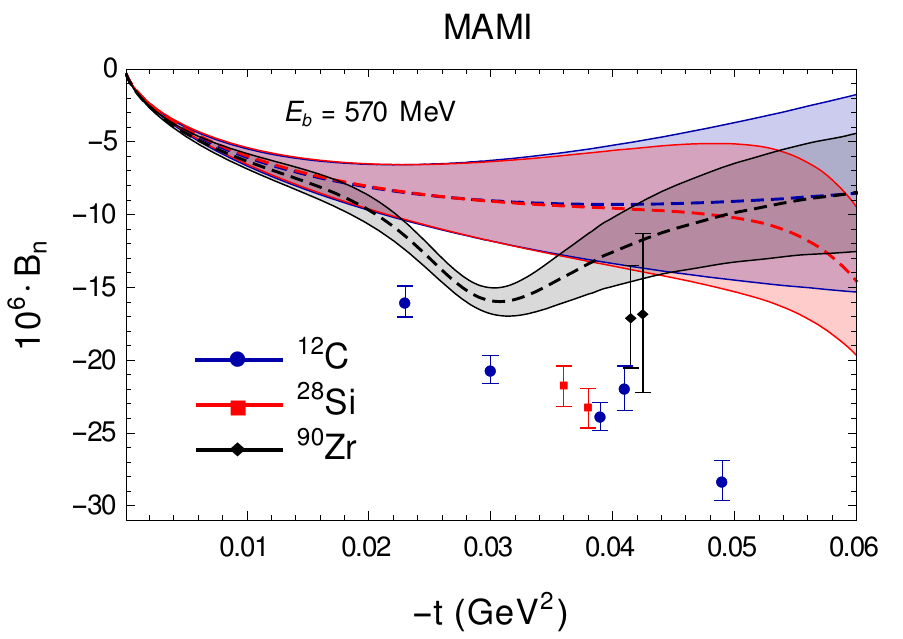}
  \caption
  {
  BNSSA versus momentum transfer squared $|t|$ in the
  kinematical range where measurements are available.
  Left panel: Predictions for $^{4}$He obtained with
  $E_{\rm b} = 2.750$~GeV and for $^{12}$C and $^{208}$Pb
  with $E_{\rm b} = 1.063$~GeV. Experimental data points are
  from the PREX-I and HAPPEX experiments \cite{AbrahamyanBnSSA2012}.
  Right panel: Predictions for $^{12}$C, $^{28}$Si, and
  $^{90}$Zr obtained with $E_{\rm b} = 570$~MeV. Experimental
  data points are from the experiments of
  Refs.~\cite{EsserBNSSA2018, Esser:2020vjb} at MAMI.
  }
  \label{fig:PREX-I+Mainz}
\end{figure*}

\section{Results}
\label{Sec:results}

In this section, we present results for $B_n$ for electron
scattering at energies ranging from 570~MeV to 3~GeV
from a variety of nuclear targets. We note here that while our 
formalism was developed for spin-0 nuclei, 
for elastic scattering on an unpolarized nuclear target non-zero 
nuclear spin will only induce corrections of the order of the nuclear recoil, $\sim t/M^2$,
which can be safely neglected. 

The results of the calculation including Coulomb distortion
(distorted-wave calculation, for short) of the
beam-normal SSA for the case when only elastic
intermediate-state contributions in the scattering process
are taken into account ($V_{\rm abs} = 0$) are displayed
in Fig.~\ref{fig:SSA_no_vabs}. This figure illustrates the
dependence of the asymmetry on details of the nuclear charge
distribution at a fixed energy of the incoming beam,
$E_{\rm b} = 953$~MeV. Because the nuclear
charge density is roughly represented by a nearly homogeneous
sphere with a relatively sharp edge, the prediction for the
beam-normal SSA features a typical diffractive pattern. The
location of the first diffraction minimum gives an idea of
the characteristic size of the target nucleus. One can see
that for light nuclei the diffraction minima are prominent
and deep, with the absolute value of the asymmetry changing
by an order of magnitude in the vicinity of the minimum.
For heavy nuclei, Coulomb distortions are stronger,
and the asymmetry experiences a less drastic change around
the minimum. The predictions for $B_n$ presented in
Fig.~\ref{fig:SSA_no_vabs} are in good agreement with
those reported in Ref.~\cite{CooperCoulDist2005}.

Next we discuss results of the distorted-wave calculation of
the beam-normal SSA for the case when inelastic intermediate-state
contributions in the scattering process are taken into account
by including the absorptive potential into the Coulomb problem.
We calculate a theoretical uncertainty in several steps. First,
we evaluate a relative uncertainty $\epsilon_1$ of the asymmetry
due to the uncertainty of $B$. The uncertainty of $B$ receives
itself two contributions: (i) the first component is the
uncertainty from the fit to the Compton data and is provided
in Table~\ref{tab:slope}; (ii) the second component is associated
with neglecting the effect of the real part of the amplitude
${\cal{F}}_3$ in the fit of the Compton data and was estimated
to be $2.2$ GeV$^{-2}$, as discussed in Sec.~\ref{Sec:abs_tpe}.
These two parts are combined in quadrature. Second, we evaluate
a contribution $\epsilon_2$ to the relative uncertainty of the
asymmetry due to the specific choice of an ansatz for the
$t$-dependence of the coefficient ${\cal{A}}_1^{(1)}(t)$.
$\epsilon_2$ is obtained as the relative difference between
predictions for $B_n$ computed with and without the contribution
from ${\cal{A}}_1^{(1)}(t)$ to $V_{\rm abs}$, while the parameter
$B$ is kept fixed at its central value. This prescription is
equivalent to assigning a 100\% uncertainty to the contribution
from ${\cal{A}}_1^{(1)}$. Finally, the two components are added
in quadrature, i.e.\ $\epsilon=\sqrt{\epsilon_1^2+\epsilon_2^2}$
is used to calculate uncertainty bands shown in the following
figures.

In Figs.~\ref{fig:PREX-I+Mainz} and \ref{fig:PREX-II+CREX}, we
display results for the BNSSA in the distorted-wave calculation
including inelastic intermediate-states. Each curve in these
figures belongs to a specific energy of the incoming beam as
specified in the figure captions and a specific target nucleus
as indicated on the plots. The central dashed lines correspond
to the absorptive potential given by
Eq.~(\ref{Eq:abs_potential_final}) and the parameter $B$ fixed
at its central value as provided in Table~\ref{tab:slope}
(5 GeV data). The solid bands around the central lines indicate
the estimated theoretical uncertainty as described in the
previous paragraph.
By comparing the results presented in
Fig.~\ref{fig:SSA_no_vabs} with those displayed on the left panel
of Fig.~\ref{fig:PREX-II+CREX} (both figures correspond to
$E_{\rm b} = 953$~MeV), we conclude that the inelastic excitations
of the intermediate state provide the dominant contribution to
$B_n$ at GeV beam energies. This is consistent with the results
of Ref.~\cite{GorsteinBNSSA2008}, in which only the leading-order
inelastic intermediate-state excitations were considered.

In Fig.~\ref{fig:PREX-I+Mainz}, we compare our prediction for
$B_n$ with the measurements by the PREX-I and HAPPEX collaborations at
JLab \cite{AbrahamyanBnSSA2012} (left plot) and a series of experiments
performed at MAMI \cite{EsserBNSSA2018, Esser:2020vjb} (right plot). 
We note that our framework has been designed for high-energy electron scattering; 
apart from lacking contributions from the nuclear range, it operates with a phenomenological 
$t$-dependence motivated by the high-energy Compton scattering data. 
While the high-energy measurement on $^4$He by the HAPPEX collaboration at $2.75$\,GeV is well described, 
and so is a somewhat lower one on $^{12}$C at $1.063$\,GeV, 
the agreement at lower MAMI energies is worse even for light and intermediate nuclei. 
This fact indicates that the $t$-dependence of the Compton cross section in the resonance region is 
likely not to follow the exponential fall-off as deduced from high-energy data. 

The data point by the PREX-I collaboration on the $^{208}$Pb target  clearly stands out: the measured value of $B_n\approx +0.5$\,p.p.m. does 
not follow the pattern of either the theoretical predictions, nor measurements on lighter nuclei, with large negative asymmetries, 
hence the name ``the PREX puzzle".
Although the distorted-wave calculation of $B_n$
reported here and obtained with the updated value
of the slope parameter $B$ reduces the 
disagreement between theory and experiment for $^{208}$Pb somewhat, it
is still unable to explain the origin of the sign difference
between measurement and prediction.

We note that the
predictions displayed in Fig.~\ref{fig:PREX-I+Mainz} are
obtained using different values of the parameter $B$ (see
Table~\ref{tab:slope} for details) for different nuclei. These
values were deduced from the Compton scattering data on 8 nuclei
\cite{Criegee:1977uf,Aleksanian:1986hb}. In contrast, theoretical
predictions presented in
Refs.~\cite{AbrahamyanBnSSA2012,EsserBNSSA2018, Esser:2020vjb}
were based on the calculation of Ref.~\cite{GorsteinBNSSA2008}
which assumed a universal parameter $B = 8\pm1$~GeV$^{-2}$, independent of
the target nucleus.
{
This value stems from the high-energy Compton data on the
proton. In Ref.~\cite{GorsteinBNSSA2008} this value was found
consistent with that for $^4$He, thereby conjecturing that it
remains constant across the nuclear chart. The present, more
careful study addressed the validity of this assumption explicitly,
see Table~\ref{tab:slope}, and found it to hold for light nuclei,
from $^4$He to $^{27}$Al. For heavier nuclei it gradually breaks
down and for the heaviest nucleus, $^{197}$Au the actual value of
the slope is 7 times larger.

In Refs.~\cite{EsserBNSSA2018, Esser:2020vjb}, light and
intermediate nuclei had been studied at lower energies. In
those references, the slope $B$ was taken universal
and constant \cite{GorsteinBNSSA2008}, but the uncertainty
was assumed to be 10\% (20\%) of the full slope of the Compton
cross section, i.e.\ of $B+R_{\rm Ch}^2/3$, with the nuclear
charge radius $R_{\rm Ch}$. For
carbon, one has $R_{\rm Ch}\approx 2.5$~fm which leads to
$B = 8 \pm 6~(\pm12)$~GeV$^{-2}$ for 10\% (20\%) uncertainty,
respectively. This is a conservative estimate in view of
experimental data that allow us to reduce the uncertainty of
$B$ considerably, as shown in Table~\ref{tab:slope}.

Another difference between the approach of
Refs.~\cite{EsserBNSSA2018, Esser:2020vjb} and the one used
in the present work concerns the treatment of corrections to
the leading-order behavior of the potential $V_{\rm abs}$
and related uncertainties. In those references,
the approximate result for $I_0$ shown in
Eq.~(\ref{eq:MasterIntegrals}) was used to obtain the central
value, while the $t$-independent non-logarithmic term
appearing in the full result of Eqs.~(\ref{Eq:feynman_par_b},
\ref{Eq:master_int_b}), was only used to estimate the uncertainty. 
Here we argue that the full expression for
$V_{\rm abs}^{(0)}$ and the subleading contribution
$V_{\rm abs}^{(1)}$ are exactly calculable and should therefore
be included in the central value. The leading term is
model-independent as it is the only term that carries the
long-range behavior $\sim\ln(|t|/m^2)$. All other corrections,
including $V_{\rm abs}^{(1)}$, are of short-range nature. Among
these, $V_{\rm abs}^{(1)}$ is the only
term enhanced by the collinear logarithm $\sim\ln(4E^2/m^2)$,
and its coefficient is exactly calculable, based on the
low-$Q^2$ expansion of the near-forward virtual Compton
amplitude. This enhancement justifies using 100\% of this
contribution as a conservative uncertainty estimate for all
neglected short-range pieces.

\begin{figure*}[t]
  \includegraphics[scale=0.9]{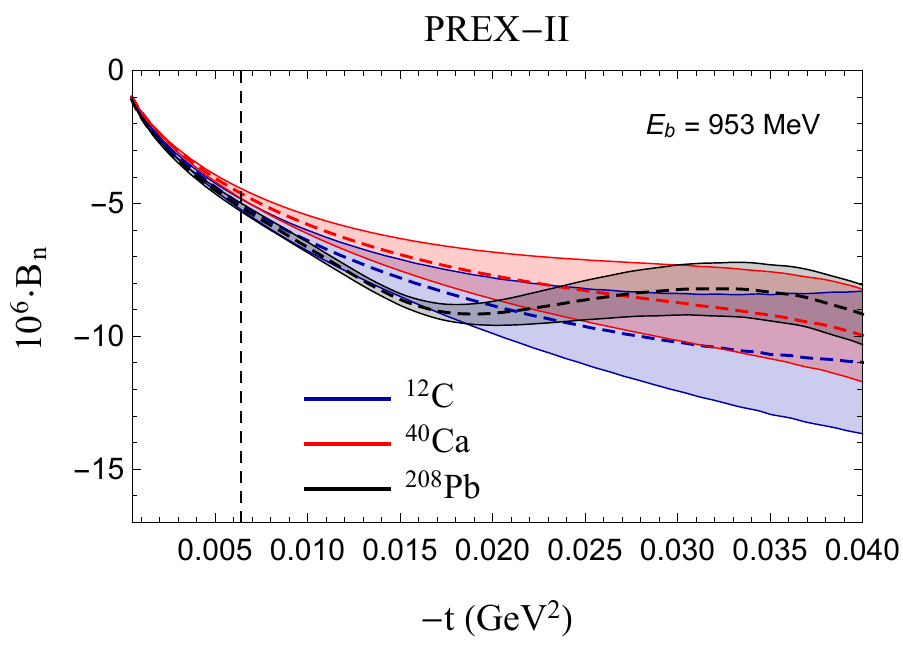}
  \includegraphics[scale=0.9]{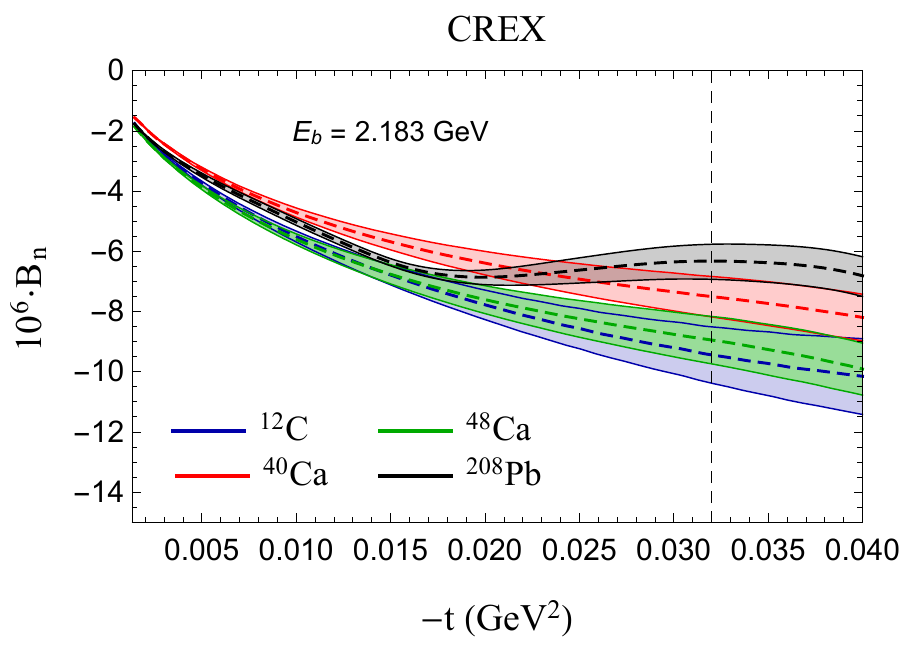}
    \includegraphics[scale=0.9]{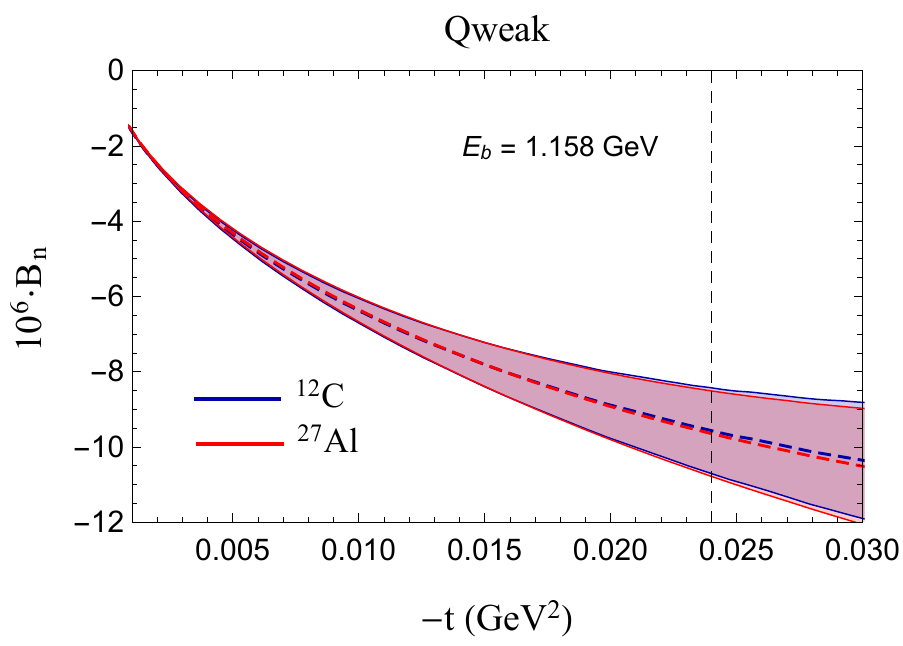}
  \caption
  {
  BNSSA versus momentum transfer squared $|t|$ for the
  kinematical conditions of soon-to-be published measurements.
  Upper left panel: Predictions for $^{12}$C, $^{40}$Ca, and
  $^{208}$Pb obtained for $E_{\rm b} = 953$~MeV (the PREX-II 
  measurement \cite{PREX-II}).
  Upper right panel: Predictions for $^{12}$C, $^{40}$Ca, 
  $^{48}$Ca, and $^{208}$Pb obtained for $E_{\rm b} = 2.183$~GeV
  (the CREX measurement \cite{CREX}). 
  Lower panel: Predictions for $^{12}$C and $^{27}$Al
  obtained for $E_{\rm b} = 1.158$~GeV (the Qweak measurement). 
  The dashed vertical lines indicate 
  approximate values of $t$ of the considered experiments.
  }
  \label{fig:PREX-II+CREX}
\end{figure*}

The numerical hierarchy of parts of $V_{\rm abs}$ may break down upon
including effects originating from low-energy nuclear excitations
and the quasielastic peak. Such contributions carry
a new intermediate scale $\Lambda_{\rm Nucl}\sim15$\,MeV with $m \ll
\Lambda_{\rm Nucl} \ll E_b \sim 1$~GeV.
However, as long as the beam energy is large enough ($\gtrsim300$\,MeV) the leading term $V_{\rm abs}^{(0)}$ is exempt
from a substantial modification by the contributions from such
low energies. To see this we may use the approximate scaling of
the integrated nuclear cross section without energy weighting,
$\int \sigma_{\rm Nucl}(\omega) d\omega = \frac{NZ}{A}
\frac{\alpha}{M}$ \cite{Berman:1975tt}, with $N$ the number of
neutrons. The expected energy-weighted
result then reads, assuming $N=Z$, $\frac{1}{A} \int_0^{30\,\rm{MeV}} \omega
\sigma_{\rm Nucl}(\omega) d\omega \sim
\frac{\alpha\Lambda_{Nucl}}{4M} \lesssim 10^{-4}$. This is to be
compared with the energy-weighted integral over the hadronic
range for which we find $\int_{\omega_\pi}^{E_b}\omega \sigma_{\gamma p}(\omega)
d\omega \sim 0.3$, for $E_b=1$\,GeV and using the parametrization of
Ref.~\cite{ChristyFit2010}. 
The effect of these nuclear contributions on 
the subleading term $V_{\rm abs}^{(1)}$ remains a question
which we plan to address in a future work.
}

In Fig.~\ref{fig:PREX-II+CREX}, we present the prediction for the kinematical conditions of soon-to-be published 
measurements of $B_n$ by the 
PREX-II \cite{Adhikari:2021phr}, CREX \cite{Kumar:2020ejz}, and Qweak  \cite{Qweak_Bn} collaborations 
at Jefferson Lab.

Finally, we confront the results of our distorted-wave calculation
to those obtained in the plane-wave approximation as reported in
Ref.~\cite{GorsteinBNSSA2008}. Here, the BNSSA is given by
\begin{equation}
\label{Eq:bnssa_leadorder}
\begin{split}
    B_n = - \frac{2m \sqrt{|t|}}{\sqrt{(s-M^2)^2-s|t|}}
    \frac{\mathrm{Im}A_{1}^{2 \gamma}}{A_{2}^{1 \gamma}}
\end{split}
\end{equation}
in terms of the invariant amplitudes $A_1^{2 \gamma}$
(Eq.~(\ref{Eq:ImA12gamma})) and $A_2^{1 \gamma}$
(Eq.~(\ref{Eq:amplitudes_ope})). In order to perform a meaningful
comparison between the two calculations, we tuned the input
parameters of the plane-wave approach to be identical to the
input we used for our distorted-wave calculation, i.e.\ instead
of assuming a flat $A/Z$ dependence of the Compton slope
parameter as in Ref.~\cite{GorsteinBNSSA2008}, we used the
experimental information on its $Z(A)$ dependence as summarized
in Table~\ref{tab:slope}. In addition, instead of evaluating the
asymmetry in the leading logarithm approximation, i.e.\ by
considering only those contributions to $B_n$ coming from
approximating ${\cal{A}}_1(t)$ with ${\cal{A}}_1^{(0)}(t)$,
Eq.~(\ref{Eq:H0}), we also took into account the
${\cal{A}}_1^{(1)}(t)$ contribution to ${\cal{A}}(t)$,
Eq.~(\ref{Eq:H1}). The results of the comparison are displayed
in Fig.~\ref{fig:Comparison-CREX}. We observe that Coulomb 
distortion increases the absolute value of the asymmetry. 
While the effect (the difference between the solid and dashed curves 
of the same color) may be significant, the corresponding predictions are 
qualitatively similar to those obtained in the plane-wave approximation.

\begin{figure}[t]
  \includegraphics[scale=0.9]{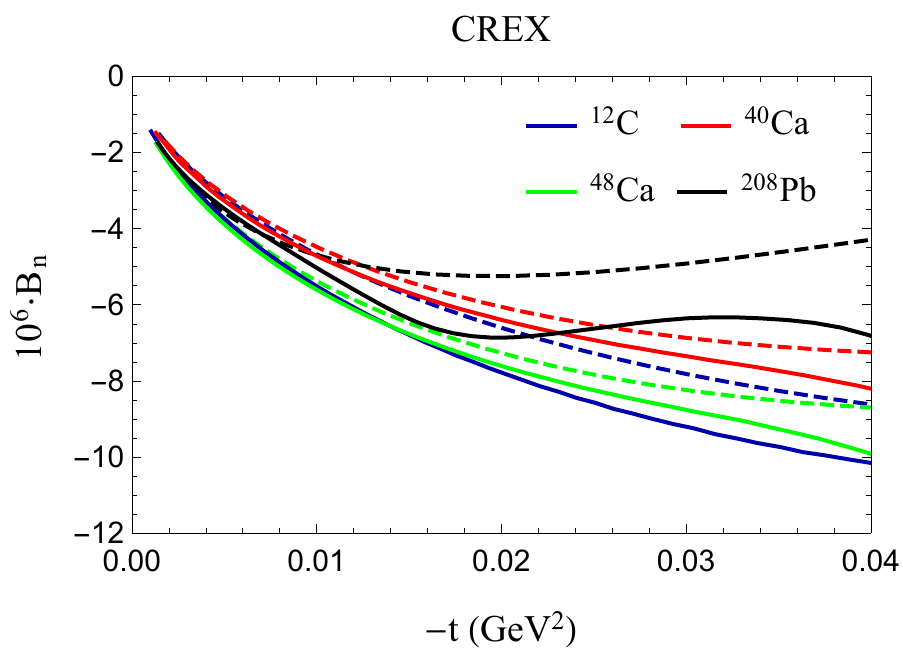}
  \caption
  {
  BNSSA versus momentum transfer squared $|t|$ for the
  kinematical conditions of the CREX measurement, i.e.\ for
  $E_{\rm b} = 2.183$~GeV. Dashed curves represent predictions
  for the asymmetry obtained in the plane-wave approximation
  while the solid curves correspond to the exact calculation
  including Coulomb distortion.
  }
  \label{fig:Comparison-CREX}
\end{figure}

\section{Conclusions}

We have computed the beam-normal single-spin asymmetry in the
diffractive regime of elastic scattering of electrons from
a variety of spin-0 nuclei. This asymmetry is generated by
the imaginary part of the interference between the direct,
$f(\theta)$, and spin-flip, $g(\theta)$, scattering amplitudes.
We have evaluated these amplitudes by studying the asymptotic
behavior of the solution of the relativistic Dirac equation
at large distances. To realistically describe the effective
interaction between the electron and target nucleus, we have
employed an optical potential method. Within this approach,
the electron-nucleus interaction is represented by
two components of the potential: the (real) Coulomb one and
(imaginary) absorptive one. The Coulomb component accounts
for the contribution from elastic intermediate states in the
scattering process, whereas the absorptive component
describes inelastic contributions. To parametrize the
absorptive component of the potential, we made use of
the result for the imaginary part of the general amplitude
$A_1$ calculated to order $\alpha^2$ in the electromagnetic
coupling. The corresponding perturbative calculation has been
performed for the kinematics of diffractive electron
scattering, where the optical theorem can be used to relate
the imaginary part of the amplitude $A_1$ to the total
photoabsorption cross section of the nucleus. To describe
the $t$-dependence of the asymmetry near the forward
scattering limit ($t\!=\!0$), we utilized information on
the $t$-dependence of the differential cross section of
Compton scattering off nuclei at low $t$. Using this
approach, we have obtained distorted-wave predictions
for BNSSA for various spin-0 nuclei and presented
the results in the kinematical range of several experiments that
have been performed. 

Our calculation contains several improvements with respect to
earlier calculations: a) we included contributions from inelastic
intermediate states into the Coulomb problem; b) we went beyond
the leading logarithm approximation in evaluating $B_n$ by
considering contributions coming from ${\cal{A}}_1^{(1)}(t)$;
c) we explicitly outlined the approximation scheme used for the
evaluation of $B_n$ and obtained a more realistic estimate
of uncertainties. 
We found however that neither of these improvements seems to be enough to 
explain the PREX puzzle. 
A small and positive value of $B_n$ obtained on a $^{208}$Pb target 
is at variance with large negative values of $B_n$ predicted by the theory and backed 
by all other measurements on light and intermediate nuclei. 
As a possible improvement, we plan to
study contributions coming from the nuclear region of
the photoabsorption cross section.

\acknowledgments
The authors acknowledge useful discussions with Chuck Horowitz, Jens Erler and
Stefan Wezorke. The work of M.~G., O.~K., and H.~S.\ is supported
by the German-Mexican research collaboration grant No.\ 278017
(CONACyT) and No.\ SP 778/4-1 (DFG). M.~G.\ is supported by the
EU Horizon 2020 research and innovation programme, project
STRONG-2020, grant agreement No.\ 824093. X.~R.-M.\ acknowledges
funding from the EU Horizon 2020 research and innovation programme,
grant agreement No.\ 654002.

\appendix

\section{Master Integrals}
\label{App:master_int}

In this Appendix, we briefly summarize the details of the
calculation of the master integrals appearing in the expression
for the beam-normal SSA, Eq.~(\ref{eq:MasterIntegrals}),
\begin{align}
\label{Eq:master_int}
    & I_0
    = \frac{|t|\vec K^2}{2\pi}\int \frac{d\Omega_K}{Q_1^2 Q_2^2}
    \, ,
    \\
    & I_1
    = \frac{E|\vec K|}{\pi}\int \frac{d\Omega_K}{Q_1^2}
    = \ln\frac{4E^2E_K^2}{m^2E_\gamma^2} \, .
\end{align}
where $E = (s - M^2)/2\sqrt{s}$ and $E_K = (s - W^2)/2\sqrt{s}$
are the energies of the external and the intermediate electrons
in the center-of-mass frame. The center-of-mass energy of the
collinear quasi-real photon is $E_\gamma = E - E_K =
(W^2 - M^2)/2\sqrt{s}$. The angular integration in
Eq.~(\ref{Eq:master_int}) can be performed by using the Feynman
trick,
\begin{align}
\label{Eq:feynman_par}
    \int \frac{d \Omega_K}{Q_1^2 Q_2^2} = \int_{0}^{1} dx
    \int \frac{d \cos \theta_x \, d \varphi}
    {\left[Q_1^2 + (Q_2^2-Q_1^2) x \right]^2} \, ,
\end{align}
and choosing the polar axis to be oriented in such a
way that $\theta_x$ is the angle between $\vec{K}$ and
$\vec{K}_x = \vec{k}_1 + x (\vec{k}_2 - \vec{k}_1)$.
Here $\vec{k}_1$ and $\vec{k}_2$ are the three-momenta
of the initial and final electron in the center-of-mass
frame. As a result, one finds \cite{AfanasevBNSSA2004,
GorsteinBNSSA2006}
\begin{align}
\label{Eq:feynman_par_b}
    I_0
    &  =
    \frac{1}{\sqrt{1+\frac{4 m^2E_\gamma^2}{|t|\vec K^2}}}
    \ln
    \frac{\sqrt{1+\frac{4 m^2E_\gamma^2}{|t|\vec K^2}}+1}%
         {\sqrt{1+\frac{4 m^2E_\gamma^2}{|t|\vec K^2}}-1} \, ,
\end{align}
where we neglected terms $\sim m^2/E^2$. In the limit
$m^2 \ll \vec K^2$, Eq.~(\ref{Eq:feynman_par_b}) reduces to
\begin{align}
\label{Eq:master_int_b}
    & I_0 = \ln\frac{|t|(E-E_\gamma)^2}{m^2E_\gamma^2} \, .
\end{align}


\begin{thebibliography}{67}%
\makeatletter
\providecommand \@ifxundefined [1]{%
 \@ifx{#1\undefined}
}%
\providecommand \@ifnum [1]{%
 \ifnum #1\expandafter \@firstoftwo
 \else \expandafter \@secondoftwo
 \fi
}%
\providecommand \@ifx [1]{%
 \ifx #1\expandafter \@firstoftwo
 \else \expandafter \@secondoftwo
 \fi
}%
\providecommand \natexlab [1]{#1}%
\providecommand \enquote  [1]{``#1''}%
\providecommand \bibnamefont  [1]{#1}%
\providecommand \bibfnamefont [1]{#1}%
\providecommand \citenamefont [1]{#1}%
\providecommand \href@noop [0]{\@secondoftwo}%
\providecommand \href [0]{\begingroup \@sanitize@url \@href}%
\providecommand \@href[1]{\@@startlink{#1}\@@href}%
\providecommand \@@href[1]{\endgroup#1\@@endlink}%
\providecommand \@sanitize@url [0]{\catcode `\\12\catcode `\$12\catcode
  `\&12\catcode `\#12\catcode `\^12\catcode `\_12\catcode `\%12\relax}%
\providecommand \@@startlink[1]{}%
\providecommand \@@endlink[0]{}%
\providecommand \url  [0]{\begingroup\@sanitize@url \@url }%
\providecommand \@url [1]{\endgroup\@href {#1}{\urlprefix }}%
\providecommand \urlprefix  [0]{URL }%
\providecommand \Eprint [0]{\href }%
\providecommand \doibase [0]{http://dx.doi.org/}%
\providecommand \selectlanguage [0]{\@gobble}%
\providecommand \bibinfo  [0]{\@secondoftwo}%
\providecommand \bibfield  [0]{\@secondoftwo}%
\providecommand \translation [1]{[#1]}%
\providecommand \BibitemOpen [0]{}%
\providecommand \bibitemStop [0]{}%
\providecommand \bibitemNoStop [0]{.\EOS\space}%
\providecommand \EOS [0]{\spacefactor3000\relax}%
\providecommand \BibitemShut  [1]{\csname bibitem#1\endcsname}%
\let\auto@bib@innerbib\@empty
\bibitem [{\citenamefont {Carlson}\ and\ \citenamefont
  {Vanderhaeghen}(2007)}]{CarlsonTPE2007}%
  \BibitemOpen
  \bibfield  {author} {\bibinfo {author} {\bibfnamefont {C.~E.}\ \bibnamefont
  {Carlson}}\ and\ \bibinfo {author} {\bibfnamefont {M.}~\bibnamefont
  {Vanderhaeghen}},\ }\href {\doibase 10.1146/annurev.nucl.57.090506.123116}
  {\bibfield  {journal} {\bibinfo  {journal} {Annu. Rev. Nucl. Part. Sci.}\
  }\textbf {\bibinfo {volume} {57}},\ \bibinfo {pages} {171} (\bibinfo {year}
  {2007})}\BibitemShut {NoStop}%
\bibitem [{\citenamefont {Arrington}\ \emph {et~al.}(2011)\citenamefont
  {Arrington}, \citenamefont {Blunden},\ and\ \citenamefont
  {Melnitchouk}}]{ArringtonTPE2011}%
  \BibitemOpen
  \bibfield  {author} {\bibinfo {author} {\bibfnamefont {J.}~\bibnamefont
  {Arrington}}, \bibinfo {author} {\bibfnamefont {P.}~\bibnamefont {Blunden}},
  \ and\ \bibinfo {author} {\bibfnamefont {W.}~\bibnamefont {Melnitchouk}},\
  }\href {\doibase 10.1016/j.ppnp.2011.07.003} {\bibfield  {journal} {\bibinfo
  {journal} {Prog. Part. Nucl. Phys.}\ }\textbf {\bibinfo {volume} {66}},\
  \bibinfo {pages} {782 } (\bibinfo {year} {2011})}\BibitemShut {NoStop}%
\bibitem [{\citenamefont {Gorchtein}(2014)}]{GorchteinTPE2014}%
  \BibitemOpen
  \bibfield  {author} {\bibinfo {author} {\bibfnamefont {M.}~\bibnamefont
  {Gorchtein}},\ }\href {\doibase 10.1103/PhysRevC.90.052201} {\bibfield
  {journal} {\bibinfo  {journal} {Phys. Rev. C}\ }\textbf {\bibinfo {volume}
  {90}},\ \bibinfo {pages} {052201} (\bibinfo {year} {2014})}\BibitemShut
  {NoStop}%
\bibitem [{\citenamefont {Afanasev}\ \emph {et~al.}(2017)\citenamefont
  {Afanasev}, \citenamefont {Blunden}, \citenamefont {Hasell},\ and\
  \citenamefont {Raue}}]{AfanasevTPE2017}%
  \BibitemOpen
  \bibfield  {author} {\bibinfo {author} {\bibfnamefont {A.}~\bibnamefont
  {Afanasev}}, \bibinfo {author} {\bibfnamefont {P.}~\bibnamefont {Blunden}},
  \bibinfo {author} {\bibfnamefont {D.}~\bibnamefont {Hasell}}, \ and\ \bibinfo
  {author} {\bibfnamefont {B.}~\bibnamefont {Raue}},\ }\href {\doibase
  10.1016/j.ppnp.2017.03.004} {\bibfield  {journal} {\bibinfo  {journal} {Prog.
  Part. Nucl. Phys.}\ }\textbf {\bibinfo {volume} {95}},\ \bibinfo {pages} {245
  } (\bibinfo {year} {2017})}\BibitemShut {NoStop}%
\bibitem [{\citenamefont {Koshchii}\ and\ \citenamefont
  {Afanasev}(2017)}]{KoshchiiAsymmetry2017}%
  \BibitemOpen
  \bibfield  {author} {\bibinfo {author} {\bibfnamefont {O.}~\bibnamefont
  {Koshchii}}\ and\ \bibinfo {author} {\bibfnamefont {A.}~\bibnamefont
  {Afanasev}},\ }\href {\doibase 10.1103/PhysRevD.96.016005} {\bibfield
  {journal} {\bibinfo  {journal} {Phys. Rev. D}\ }\textbf {\bibinfo {volume}
  {96}},\ \bibinfo {pages} {016005} (\bibinfo {year} {2017})}\BibitemShut
  {NoStop}%
\bibitem [{\citenamefont {Talukdar}\ \emph {et~al.}(2020)\citenamefont
  {Talukdar}, \citenamefont {Shastry}, \citenamefont {Raha},\ and\
  \citenamefont {Myhrer}}]{Pulak.ChiralTPE.2020}%
  \BibitemOpen
  \bibfield  {author} {\bibinfo {author} {\bibfnamefont {P.}~\bibnamefont
  {Talukdar}}, \bibinfo {author} {\bibfnamefont {V.~C.}\ \bibnamefont
  {Shastry}}, \bibinfo {author} {\bibfnamefont {U.}~\bibnamefont {Raha}}, \
  and\ \bibinfo {author} {\bibfnamefont {F.}~\bibnamefont {Myhrer}},\ }\href
  {\doibase 10.1103/PhysRevD.101.013008} {\bibfield  {journal} {\bibinfo
  {journal} {Phys. Rev. D}\ }\textbf {\bibinfo {volume} {101}},\ \bibinfo
  {pages} {013008} (\bibinfo {year} {2020})}\BibitemShut {NoStop}%
\bibitem [{\citenamefont {Cao}\ and\ \citenamefont {Zhou}(2020)}]{ZhouTPE2020}%
  \BibitemOpen
  \bibfield  {author} {\bibinfo {author} {\bibfnamefont {H.-Y.}\ \bibnamefont
  {Cao}}\ and\ \bibinfo {author} {\bibfnamefont {H.-Q.}\ \bibnamefont {Zhou}},\
  }\href@noop {} {\  (\bibinfo {year} {2020})},\ \Eprint
  {http://arxiv.org/abs/2005.08265} {arXiv:2005.08265 [nucl-th]} \BibitemShut
  {NoStop}%
\bibitem [{\citenamefont {Rachek}\ \emph {et~al.}(2015)\citenamefont {Rachek}
  \emph {et~al.}}]{RachekTPE2015}%
  \BibitemOpen
  \bibfield  {author} {\bibinfo {author} {\bibfnamefont {I.~A.}\ \bibnamefont
  {Rachek}} \emph {et~al.},\ }\href {\doibase 10.1103/PhysRevLett.114.062005}
  {\bibfield  {journal} {\bibinfo  {journal} {Phys. Rev. Lett.}\ }\textbf
  {\bibinfo {volume} {114}},\ \bibinfo {pages} {062005} (\bibinfo {year}
  {2015})}\BibitemShut {NoStop}%
\bibitem [{\citenamefont {Rimal}\ \emph {et~al.}(2017)\citenamefont {Rimal}
  \emph {et~al.}}]{RimalTPE2017}%
  \BibitemOpen
  \bibfield  {author} {\bibinfo {author} {\bibfnamefont {D.}~\bibnamefont
  {Rimal}} \emph {et~al.} (\bibinfo {collaboration} {CLAS Collaboration}),\
  }\href {\doibase 10.1103/PhysRevC.95.065201} {\bibfield  {journal} {\bibinfo
  {journal} {Phys. Rev. C}\ }\textbf {\bibinfo {volume} {95}},\ \bibinfo
  {pages} {065201} (\bibinfo {year} {2017})}\BibitemShut {NoStop}%
\bibitem [{\citenamefont {Henderson}\ \emph {et~al.}(2017)\citenamefont
  {Henderson} \emph {et~al.}}]{HendersonTPEOlympus2017}%
  \BibitemOpen
  \bibfield  {author} {\bibinfo {author} {\bibfnamefont {B.~S.}\ \bibnamefont
  {Henderson}} \emph {et~al.} (\bibinfo {collaboration} {OLYMPUS
  Collaboration}),\ }\href {\doibase 10.1103/PhysRevLett.118.092501} {\bibfield
   {journal} {\bibinfo  {journal} {Phys. Rev. Lett.}\ }\textbf {\bibinfo
  {volume} {118}},\ \bibinfo {pages} {092501} (\bibinfo {year}
  {2017})}\BibitemShut {NoStop}%
\bibitem [{\citenamefont {Qattan}\ \emph {et~al.}(2018)\citenamefont {Qattan},
  \citenamefont {Homouz},\ and\ \citenamefont {Riahi}}]{QattanTPE2018}%
  \BibitemOpen
  \bibfield  {author} {\bibinfo {author} {\bibfnamefont {I.~A.}\ \bibnamefont
  {Qattan}}, \bibinfo {author} {\bibfnamefont {D.}~\bibnamefont {Homouz}}, \
  and\ \bibinfo {author} {\bibfnamefont {M.~K.}\ \bibnamefont {Riahi}},\ }\href
  {\doibase 10.1103/PhysRevC.97.045201} {\bibfield  {journal} {\bibinfo
  {journal} {Phys. Rev. C}\ }\textbf {\bibinfo {volume} {97}},\ \bibinfo
  {pages} {045201} (\bibinfo {year} {2018})}\BibitemShut {NoStop}%
\bibitem [{\citenamefont {Qattan}\ \emph {et~al.}(2020)\citenamefont {Qattan},
  \citenamefont {Patole},\ and\ \citenamefont {Alsaad}}]{QattanTPE2020}%
  \BibitemOpen
  \bibfield  {author} {\bibinfo {author} {\bibfnamefont {I.~A.}\ \bibnamefont
  {Qattan}}, \bibinfo {author} {\bibfnamefont {S.~P.}\ \bibnamefont {Patole}},
  \ and\ \bibinfo {author} {\bibfnamefont {A.}~\bibnamefont {Alsaad}},\ }\href
  {\doibase 10.1103/PhysRevC.101.055202} {\bibfield  {journal} {\bibinfo
  {journal} {Phys. Rev. C}\ }\textbf {\bibinfo {volume} {101}},\ \bibinfo
  {pages} {055202} (\bibinfo {year} {2020})}\BibitemShut {NoStop}%
\bibitem [{\citenamefont {Jones}\ \emph {et~al.}(2000)\citenamefont {Jones}
  \emph {et~al.}}]{JonesFF2000}%
  \BibitemOpen
  \bibfield  {author} {\bibinfo {author} {\bibfnamefont {M.~K.}\ \bibnamefont
  {Jones}} \emph {et~al.} (\bibinfo {collaboration} {Jefferson Lab Hall A
  Collaboration}),\ }\href {\doibase 10.1103/PhysRevLett.84.1398} {\bibfield
  {journal} {\bibinfo  {journal} {Phys. Rev. Lett.}\ }\textbf {\bibinfo
  {volume} {84}},\ \bibinfo {pages} {1398} (\bibinfo {year}
  {2000})}\BibitemShut {NoStop}%
\bibitem [{\citenamefont {Gayou}\ \emph {et~al.}(2002)\citenamefont {Gayou}
  \emph {et~al.}}]{GayouFF2002}%
  \BibitemOpen
  \bibfield  {author} {\bibinfo {author} {\bibfnamefont {O.}~\bibnamefont
  {Gayou}} \emph {et~al.} (\bibinfo {collaboration} {Jefferson Lab Hall A
  Collaboration}),\ }\href {\doibase 10.1103/PhysRevLett.88.092301} {\bibfield
  {journal} {\bibinfo  {journal} {Phys. Rev. Lett.}\ }\textbf {\bibinfo
  {volume} {88}},\ \bibinfo {pages} {092301} (\bibinfo {year}
  {2002})}\BibitemShut {NoStop}%
\bibitem [{\citenamefont {Ahmed}\ \emph {et~al.}(2020)\citenamefont {Ahmed},
  \citenamefont {Blunden},\ and\ \citenamefont {Melnitchouk}}]{AhmedTPE2020}%
  \BibitemOpen
  \bibfield  {author} {\bibinfo {author} {\bibfnamefont {J.}~\bibnamefont
  {Ahmed}}, \bibinfo {author} {\bibfnamefont {P.~G.}\ \bibnamefont {Blunden}},
  \ and\ \bibinfo {author} {\bibfnamefont {W.}~\bibnamefont {Melnitchouk}},\
  }\href {\doibase 10.1103/PhysRevC.102.045205} {\bibfield  {journal} {\bibinfo
   {journal} {Phys. Rev. C}\ }\textbf {\bibinfo {volume} {102}},\ \bibinfo
  {pages} {045205} (\bibinfo {year} {2020})}\BibitemShut {NoStop}%
\bibitem [{\citenamefont {Afanasev}\ and\ \citenamefont
  {Merenkov}(2004{\natexlab{a}})}]{AfanasevSSA2004}%
  \BibitemOpen
  \bibfield  {author} {\bibinfo {author} {\bibfnamefont {A.~V.}\ \bibnamefont
  {Afanasev}}\ and\ \bibinfo {author} {\bibfnamefont {N.~P.}\ \bibnamefont
  {Merenkov}},\ }\href {\doibase 10.1103/PhysRevD.70.073002} {\bibfield
  {journal} {\bibinfo  {journal} {Phys. Rev. D}\ }\textbf {\bibinfo {volume}
  {70}},\ \bibinfo {pages} {073002} (\bibinfo {year}
  {2004}{\natexlab{a}})}\BibitemShut {NoStop}%
\bibitem [{\citenamefont {Afanasev}\ and\ \citenamefont
  {Merenkov}(2004{\natexlab{b}})}]{AfanasevBNSSA2004}%
  \BibitemOpen
  \bibfield  {author} {\bibinfo {author} {\bibfnamefont {A.~V.}\ \bibnamefont
  {Afanasev}}\ and\ \bibinfo {author} {\bibfnamefont {N.}~\bibnamefont
  {Merenkov}},\ }\href {\doibase 10.1016/j.physletb.2004.08.023} {\bibfield
  {journal} {\bibinfo  {journal} {Phys. Lett. B}\ }\textbf {\bibinfo {volume}
  {599}},\ \bibinfo {pages} {48 } (\bibinfo {year}
  {2004}{\natexlab{b}})}\BibitemShut {NoStop}%
\bibitem [{\citenamefont {Gorchtein}(2006)}]{GorsteinBNSSA2006}%
  \BibitemOpen
  \bibfield  {author} {\bibinfo {author} {\bibfnamefont {M.}~\bibnamefont
  {Gorchtein}},\ }\href {\doibase 10.1103/PhysRevC.73.055201} {\bibfield
  {journal} {\bibinfo  {journal} {Phys. Rev. C}\ }\textbf {\bibinfo {volume}
  {73}},\ \bibinfo {pages} {055201} (\bibinfo {year} {2006})}\BibitemShut
  {NoStop}%
\bibitem [{\citenamefont {Pasquini}\ and\ \citenamefont
  {Vanderhaeghen}(2004)}]{PasquiniSSA2004}%
  \BibitemOpen
  \bibfield  {author} {\bibinfo {author} {\bibfnamefont {B.}~\bibnamefont
  {Pasquini}}\ and\ \bibinfo {author} {\bibfnamefont {M.}~\bibnamefont
  {Vanderhaeghen}},\ }\href {\doibase 10.1103/PhysRevC.70.045206} {\bibfield
  {journal} {\bibinfo  {journal} {Phys. Rev. C}\ }\textbf {\bibinfo {volume}
  {70}},\ \bibinfo {pages} {045206} (\bibinfo {year} {2004})}\BibitemShut
  {NoStop}%
\bibitem [{\citenamefont {Gorchtein}\ \emph {et~al.}(2004)\citenamefont
  {Gorchtein}, \citenamefont {Guichon},\ and\ \citenamefont
  {Vanderhaeghen}}]{GorsteinBNSSA2004}%
  \BibitemOpen
  \bibfield  {author} {\bibinfo {author} {\bibfnamefont {M.}~\bibnamefont
  {Gorchtein}}, \bibinfo {author} {\bibfnamefont {P.}~\bibnamefont {Guichon}},
  \ and\ \bibinfo {author} {\bibfnamefont {M.}~\bibnamefont {Vanderhaeghen}},\
  }\href {\doibase 10.1016/j.nuclphysa.2004.06.008} {\bibfield  {journal}
  {\bibinfo  {journal} {Nucl. Phys. A}\ }\textbf {\bibinfo {volume} {741}},\
  \bibinfo {pages} {234 } (\bibinfo {year} {2004})}\BibitemShut {NoStop}%
\bibitem [{\citenamefont {Cooper}\ and\ \citenamefont
  {Horowitz}(2005)}]{CooperCoulDist2005}%
  \BibitemOpen
  \bibfield  {author} {\bibinfo {author} {\bibfnamefont {E.~D.}\ \bibnamefont
  {Cooper}}\ and\ \bibinfo {author} {\bibfnamefont {C.~J.}\ \bibnamefont
  {Horowitz}},\ }\href {\doibase 10.1103/PhysRevC.72.034602} {\bibfield
  {journal} {\bibinfo  {journal} {Phys. Rev. C}\ }\textbf {\bibinfo {volume}
  {72}},\ \bibinfo {pages} {034602} (\bibinfo {year} {2005})}\BibitemShut
  {NoStop}%
\bibitem [{\citenamefont {Borisyuk}\ and\ \citenamefont
  {Kobushkin}(2006)}]{BorisyukBNSSA2006}%
  \BibitemOpen
  \bibfield  {author} {\bibinfo {author} {\bibfnamefont {D.}~\bibnamefont
  {Borisyuk}}\ and\ \bibinfo {author} {\bibfnamefont {A.}~\bibnamefont
  {Kobushkin}},\ }\href {\doibase 10.1103/PhysRevC.73.045210} {\bibfield
  {journal} {\bibinfo  {journal} {Phys. Rev. C}\ }\textbf {\bibinfo {volume}
  {73}},\ \bibinfo {pages} {045210} (\bibinfo {year} {2006})}\BibitemShut
  {NoStop}%
\bibitem [{\citenamefont {Gorchtein}(2007)}]{GorchteinDispersive2007}%
  \BibitemOpen
  \bibfield  {author} {\bibinfo {author} {\bibfnamefont {M.}~\bibnamefont
  {Gorchtein}},\ }\href {\doibase 10.1016/j.physletb.2006.11.065} {\bibfield
  {journal} {\bibinfo  {journal} {Phys. Lett. B}\ }\textbf {\bibinfo {volume}
  {644}},\ \bibinfo {pages} {322 } (\bibinfo {year} {2007})}\BibitemShut
  {NoStop}%
\bibitem [{\citenamefont {Gorchtein}\ and\ \citenamefont
  {Horowitz}(2008)}]{GorsteinBNSSA2008}%
  \BibitemOpen
  \bibfield  {author} {\bibinfo {author} {\bibfnamefont {M.}~\bibnamefont
  {Gorchtein}}\ and\ \bibinfo {author} {\bibfnamefont {C.~J.}\ \bibnamefont
  {Horowitz}},\ }\href {\doibase 10.1103/PhysRevC.77.044606} {\bibfield
  {journal} {\bibinfo  {journal} {Phys. Rev. C}\ }\textbf {\bibinfo {volume}
  {77}},\ \bibinfo {pages} {044606} (\bibinfo {year} {2008})}\BibitemShut
  {NoStop}%
\bibitem [{\citenamefont {Carlson}\ \emph {et~al.}(2017)\citenamefont
  {Carlson}, \citenamefont {Pasquini}, \citenamefont {Pauk},\ and\
  \citenamefont {Vanderhaeghen}}]{CarlsonSSA2017}%
  \BibitemOpen
  \bibfield  {author} {\bibinfo {author} {\bibfnamefont {C.~E.}\ \bibnamefont
  {Carlson}}, \bibinfo {author} {\bibfnamefont {B.}~\bibnamefont {Pasquini}},
  \bibinfo {author} {\bibfnamefont {V.}~\bibnamefont {Pauk}}, \ and\ \bibinfo
  {author} {\bibfnamefont {M.}~\bibnamefont {Vanderhaeghen}},\ }\href {\doibase
  10.1103/PhysRevD.96.113010} {\bibfield  {journal} {\bibinfo  {journal} {Phys.
  Rev. D}\ }\textbf {\bibinfo {volume} {96}},\ \bibinfo {pages} {113010}
  (\bibinfo {year} {2017})}\BibitemShut {NoStop}%
\bibitem [{\citenamefont {Koshchii}\ and\ \citenamefont
  {Afanasev}(2018)}]{KoshchiiTNSSAen2018}%
  \BibitemOpen
  \bibfield  {author} {\bibinfo {author} {\bibfnamefont {O.}~\bibnamefont
  {Koshchii}}\ and\ \bibinfo {author} {\bibfnamefont {A.}~\bibnamefont
  {Afanasev}},\ }\href {\doibase 10.1103/PhysRevD.98.056007} {\bibfield
  {journal} {\bibinfo  {journal} {Phys. Rev. D}\ }\textbf {\bibinfo {volume}
  {98}},\ \bibinfo {pages} {056007} (\bibinfo {year} {2018})}\BibitemShut
  {NoStop}%
\bibitem [{\citenamefont {Koshchii}\ and\ \citenamefont
  {Afanasev}(2019)}]{KoshchiiBnSSAmup2019}%
  \BibitemOpen
  \bibfield  {author} {\bibinfo {author} {\bibfnamefont {O.}~\bibnamefont
  {Koshchii}}\ and\ \bibinfo {author} {\bibfnamefont {A.}~\bibnamefont
  {Afanasev}},\ }\href {\doibase 10.1103/PhysRevD.100.096020} {\bibfield
  {journal} {\bibinfo  {journal} {Phys. Rev. D}\ }\textbf {\bibinfo {volume}
  {100}},\ \bibinfo {pages} {096020} (\bibinfo {year} {2019})}\BibitemShut
  {NoStop}%
\bibitem [{\citenamefont {Rujula}\ \emph {et~al.}(1971)\citenamefont {Rujula},
  \citenamefont {Kaplan},\ and\ \citenamefont {de~Rafael}}]{RujulaSSA1971}%
  \BibitemOpen
  \bibfield  {author} {\bibinfo {author} {\bibfnamefont {A.~D.}\ \bibnamefont
  {Rujula}}, \bibinfo {author} {\bibfnamefont {J.}~\bibnamefont {Kaplan}}, \
  and\ \bibinfo {author} {\bibfnamefont {E.}~\bibnamefont {de~Rafael}},\ }\href
  {\doibase 10.1016/0550-3213(71)90460-3} {\bibfield  {journal} {\bibinfo
  {journal} {Nucl. Phys. B}\ }\textbf {\bibinfo {volume} {35}},\ \bibinfo
  {pages} {365 } (\bibinfo {year} {1971})}\BibitemShut {NoStop}%
\bibitem [{\citenamefont {Wells}\ \emph {et~al.}(2001)\citenamefont {Wells}
  \emph {et~al.}}]{SampleBNSSA2001}%
  \BibitemOpen
  \bibfield  {author} {\bibinfo {author} {\bibfnamefont {S.~P.}\ \bibnamefont
  {Wells}} \emph {et~al.},\ }\href {\doibase 10.1103/PhysRevC.63.064001}
  {\bibfield  {journal} {\bibinfo  {journal} {Phys. Rev. C}\ }\textbf {\bibinfo
  {volume} {63}},\ \bibinfo {pages} {064001} (\bibinfo {year}
  {2001})}\BibitemShut {NoStop}%
\bibitem [{\citenamefont {Maas}\ \emph {et~al.}(2005)\citenamefont {Maas} \emph
  {et~al.}}]{A4Mass2005}%
  \BibitemOpen
  \bibfield  {author} {\bibinfo {author} {\bibfnamefont {F.~E.}\ \bibnamefont
  {Maas}} \emph {et~al.},\ }\href {\doibase 10.1103/PhysRevLett.94.082001}
  {\bibfield  {journal} {\bibinfo  {journal} {Phys. Rev. Lett.}\ }\textbf
  {\bibinfo {volume} {94}},\ \bibinfo {pages} {082001} (\bibinfo {year}
  {2005})}\BibitemShut {NoStop}%
\bibitem [{\citenamefont {Armstrong}\ \emph {et~al.}(2007)\citenamefont
  {Armstrong} \emph {et~al.}}]{G0BNSSA2007}%
  \BibitemOpen
  \bibfield  {author} {\bibinfo {author} {\bibfnamefont {D.~S.}\ \bibnamefont
  {Armstrong}} \emph {et~al.} (\bibinfo {collaboration} {G0 Collaboration}),\
  }\href {\doibase 10.1103/PhysRevLett.99.092301} {\bibfield  {journal}
  {\bibinfo  {journal} {Phys. Rev. Lett.}\ }\textbf {\bibinfo {volume} {99}},\
  \bibinfo {pages} {092301} (\bibinfo {year} {2007})}\BibitemShut {NoStop}%
\bibitem [{\citenamefont {Capozza}(2007)}]{A4Capozza2007}%
  \BibitemOpen
  \bibfield  {author} {\bibinfo {author} {\bibfnamefont {L.}~\bibnamefont
  {Capozza}},\ }\href {\doibase 10.1140/epja/i2006-10428-6} {\bibfield
  {journal} {\bibinfo  {journal} {Eur. Phys. J. A}\ }\textbf {\bibinfo {volume}
  {32}},\ \bibinfo {pages} {497} (\bibinfo {year} {2007})}\BibitemShut
  {NoStop}%
\bibitem [{\citenamefont {Androi\ifmmode~\acute{c}\else \'{c}\fi{}}\ \emph
  {et~al.}(2011)\citenamefont {Androi\ifmmode~\acute{c}\else \'{c}\fi{}} \emph
  {et~al.}}]{G0BNSSA2011}%
  \BibitemOpen
  \bibfield  {author} {\bibinfo {author} {\bibfnamefont {D.}~\bibnamefont
  {Androi\ifmmode~\acute{c}\else \'{c}\fi{}}} \emph {et~al.} (\bibinfo
  {collaboration} {G0 Collaboration}),\ }\href {\doibase
  10.1103/PhysRevLett.107.022501} {\bibfield  {journal} {\bibinfo  {journal}
  {Phys. Rev. Lett.}\ }\textbf {\bibinfo {volume} {107}},\ \bibinfo {pages}
  {022501} (\bibinfo {year} {2011})}\BibitemShut {NoStop}%
\bibitem [{\citenamefont {Abrahamyan}\ \emph
  {et~al.}(2012{\natexlab{a}})\citenamefont {Abrahamyan} \emph
  {et~al.}}]{AbrahamyanBnSSA2012}%
  \BibitemOpen
  \bibfield  {author} {\bibinfo {author} {\bibfnamefont {S.}~\bibnamefont
  {Abrahamyan}} \emph {et~al.} (\bibinfo {collaboration} {HAPPEX and PREX
  Collaborations}),\ }\href {\doibase 10.1103/PhysRevLett.109.192501}
  {\bibfield  {journal} {\bibinfo  {journal} {Phys. Rev. Lett.}\ }\textbf
  {\bibinfo {volume} {109}},\ \bibinfo {pages} {192501} (\bibinfo {year}
  {2012}{\natexlab{a}})}\BibitemShut {NoStop}%
\bibitem [{\citenamefont {Gou}\ \emph {et~al.}(2020)\citenamefont {Gou} \emph
  {et~al.}}]{GouBnSSA2020}%
  \BibitemOpen
  \bibfield  {author} {\bibinfo {author} {\bibfnamefont {B.}~\bibnamefont
  {Gou}} \emph {et~al.},\ }\href {\doibase 10.1103/PhysRevLett.124.122003}
  {\bibfield  {journal} {\bibinfo  {journal} {Phys. Rev. Lett.}\ }\textbf
  {\bibinfo {volume} {124}},\ \bibinfo {pages} {122003} (\bibinfo {year}
  {2020})}\BibitemShut {NoStop}%
\bibitem [{\citenamefont {Androi\'c}\ \emph {et~al.}(2020)\citenamefont
  {Androi\'c} \emph {et~al.}}]{Androic:2020rkw}%
  \BibitemOpen
  \bibfield  {author} {\bibinfo {author} {\bibfnamefont {D.}~\bibnamefont
  {Androi\'c}} \emph {et~al.} (\bibinfo {collaboration} {QWeak}),\ }\href
  {\doibase 10.1103/PhysRevLett.125.112502} {\bibfield  {journal} {\bibinfo
  {journal} {Phys. Rev. Lett.}\ }\textbf {\bibinfo {volume} {125}},\ \bibinfo
  {pages} {112502} (\bibinfo {year} {2020})},\ \Eprint
  {http://arxiv.org/abs/2006.12435} {arXiv:2006.12435 [nucl-ex]} \BibitemShut
  {NoStop}%
\bibitem [{\citenamefont {Esser}\ \emph {et~al.}(2018)\citenamefont {Esser}
  \emph {et~al.}}]{EsserBNSSA2018}%
  \BibitemOpen
  \bibfield  {author} {\bibinfo {author} {\bibfnamefont {A.}~\bibnamefont
  {Esser}} \emph {et~al.},\ }\href {\doibase 10.1103/PhysRevLett.121.022503}
  {\bibfield  {journal} {\bibinfo  {journal} {Phys. Rev. Lett.}\ }\textbf
  {\bibinfo {volume} {121}},\ \bibinfo {pages} {022503} (\bibinfo {year}
  {2018})}\BibitemShut {NoStop}%
\bibitem [{\citenamefont {Esser}\ \emph {et~al.}(2020)\citenamefont {Esser}
  \emph {et~al.}}]{Esser:2020vjb}%
  \BibitemOpen
  \bibfield  {author} {\bibinfo {author} {\bibfnamefont {A.}~\bibnamefont
  {Esser}} \emph {et~al.},\ }\href {\doibase
  https://doi.org/10.1016/j.physletb.2020.135664} {\bibfield  {journal}
  {\bibinfo  {journal} {Phys. Lett. B}\ }\textbf {\bibinfo {volume} {808}},\
  \bibinfo {pages} {135664} (\bibinfo {year} {2020})}\BibitemShut {NoStop}%
\bibitem [{\citenamefont {Androi\'c}\ \emph {et~al.}(2013)\citenamefont
  {Androi\'c} \emph {et~al.}}]{AndroicQweak2013}%
  \BibitemOpen
  \bibfield  {author} {\bibinfo {author} {\bibfnamefont {D.}~\bibnamefont
  {Androi\'c}} \emph {et~al.} (\bibinfo {collaboration} {Qweak
  Collaboration}),\ }\href {\doibase 10.1103/PhysRevLett.111.141803} {\bibfield
   {journal} {\bibinfo  {journal} {Phys. Rev. Lett.}\ }\textbf {\bibinfo
  {volume} {111}},\ \bibinfo {pages} {141803} (\bibinfo {year}
  {2013})}\BibitemShut {NoStop}%
\bibitem [{\citenamefont {Androi\'c}\ \emph {et~al.}(2018)\citenamefont
  {Androi\'c} \emph {et~al.}}]{AndroicQWeak2018}%
  \BibitemOpen
  \bibfield  {author} {\bibinfo {author} {\bibfnamefont {D.}~\bibnamefont
  {Androi\'c}} \emph {et~al.} (\bibinfo {collaboration} {Qweak
  Collaboration}),\ }\href {\doibase 10.1038/s41586-018-0096-0} {\bibfield
  {journal} {\bibinfo  {journal} {Nature}\ }\textbf {\bibinfo {volume} {557}},\
  \bibinfo {pages} {207} (\bibinfo {year} {2018})},\ \Eprint
  {http://arxiv.org/abs/1905.08283} {arXiv:1905.08283 [nucl-ex]} \BibitemShut
  {NoStop}%
\bibitem [{\citenamefont {{Becker, Dominik}}\ \emph {et~al.}(2018)\citenamefont
  {{Becker, Dominik}} \emph {et~al.}}]{BeckerP22018}%
  \BibitemOpen
  \bibfield  {author} {\bibinfo {author} {\bibnamefont {{Becker, Dominik}}}
  \emph {et~al.},\ }\href {\doibase 10.1140/epja/i2018-12611-6} {\bibfield
  {journal} {\bibinfo  {journal} {Eur. Phys. J. A}\ }\textbf {\bibinfo {volume}
  {54}},\ \bibinfo {pages} {208} (\bibinfo {year} {2018})}\BibitemShut
  {NoStop}%
\bibitem [{\citenamefont {Horowitz}\ \emph {et~al.}(2012)\citenamefont
  {Horowitz} \emph {et~al.}}]{HorowitzPREX2012}%
  \BibitemOpen
  \bibfield  {author} {\bibinfo {author} {\bibfnamefont {C.~J.}\ \bibnamefont
  {Horowitz}} \emph {et~al.},\ }\href {\doibase 10.1103/PhysRevC.85.032501}
  {\bibfield  {journal} {\bibinfo  {journal} {Phys. Rev. C}\ }\textbf {\bibinfo
  {volume} {85}},\ \bibinfo {pages} {032501} (\bibinfo {year}
  {2012})}\BibitemShut {NoStop}%
\bibitem [{\citenamefont {Abrahamyan}\ \emph
  {et~al.}(2012{\natexlab{b}})\citenamefont {Abrahamyan} \emph
  {et~al.}}]{AbrahamyanPREX2012}%
  \BibitemOpen
  \bibfield  {author} {\bibinfo {author} {\bibfnamefont {S.}~\bibnamefont
  {Abrahamyan}} \emph {et~al.} (\bibinfo {collaboration} {PREX
  Collaboration}),\ }\href {\doibase 10.1103/PhysRevLett.108.112502} {\bibfield
   {journal} {\bibinfo  {journal} {Phys. Rev. Lett.}\ }\textbf {\bibinfo
  {volume} {108}},\ \bibinfo {pages} {112502} (\bibinfo {year}
  {2012}{\natexlab{b}})}\BibitemShut {NoStop}%
\bibitem [{\citenamefont {Yang}\ \emph {et~al.}(2019)\citenamefont {Yang},
  \citenamefont {Hernandez},\ and\ \citenamefont
  {Piekarewicz}}]{YangEWprobes2019}%
  \BibitemOpen
  \bibfield  {author} {\bibinfo {author} {\bibfnamefont {J.}~\bibnamefont
  {Yang}}, \bibinfo {author} {\bibfnamefont {J.~A.}\ \bibnamefont {Hernandez}},
  \ and\ \bibinfo {author} {\bibfnamefont {J.}~\bibnamefont {Piekarewicz}},\
  }\href {\doibase 10.1103/PhysRevC.100.054301} {\bibfield  {journal} {\bibinfo
   {journal} {Phys. Rev. C}\ }\textbf {\bibinfo {volume} {100}},\ \bibinfo
  {pages} {054301} (\bibinfo {year} {2019})}\BibitemShut {NoStop}%
\bibitem [{\citenamefont {Koshchii}\ \emph {et~al.}(2020)\citenamefont
  {Koshchii}, \citenamefont {Erler}, \citenamefont {Gorchtein}, \citenamefont
  {Horowitz}, \citenamefont {Piekarewicz}, \citenamefont {Roca-Maza},
  \citenamefont {Seng},\ and\ \citenamefont {Spiesberger}}]{Koshchii:2020qkr}%
  \BibitemOpen
  \bibfield  {author} {\bibinfo {author} {\bibfnamefont {O.}~\bibnamefont
  {Koshchii}}, \bibinfo {author} {\bibfnamefont {J.}~\bibnamefont {Erler}},
  \bibinfo {author} {\bibfnamefont {M.}~\bibnamefont {Gorchtein}}, \bibinfo
  {author} {\bibfnamefont {C.~J.}\ \bibnamefont {Horowitz}}, \bibinfo {author}
  {\bibfnamefont {J.}~\bibnamefont {Piekarewicz}}, \bibinfo {author}
  {\bibfnamefont {X.}~\bibnamefont {Roca-Maza}}, \bibinfo {author}
  {\bibfnamefont {C.-Y.}\ \bibnamefont {Seng}}, \ and\ \bibinfo {author}
  {\bibfnamefont {H.}~\bibnamefont {Spiesberger}},\ }\href {\doibase
  10.1103/PhysRevC.102.022501} {\bibfield  {journal} {\bibinfo  {journal}
  {Phys. Rev. C}\ }\textbf {\bibinfo {volume} {102}},\ \bibinfo {pages}
  {022501} (\bibinfo {year} {2020})}\BibitemShut {NoStop}%
\bibitem [{\citenamefont {Vries}\ \emph {et~al.}(1987)\citenamefont {Vries},
  \citenamefont {Jager},\ and\ \citenamefont {Vries}}]{Vries1987}%
  \BibitemOpen
  \bibfield  {author} {\bibinfo {author} {\bibfnamefont {H.~D.}\ \bibnamefont
  {Vries}}, \bibinfo {author} {\bibfnamefont {C.~D.}\ \bibnamefont {Jager}}, \
  and\ \bibinfo {author} {\bibfnamefont {C.~D.}\ \bibnamefont {Vries}},\ }\href
  {\doibase 10.1016/0092-640X(87)90013-1} {\bibfield  {journal} {\bibinfo
  {journal} {Atomic Data and Nuclear Data Tables}\ }\textbf {\bibinfo {volume}
  {36}},\ \bibinfo {pages} {495 } (\bibinfo {year} {1987})}\BibitemShut
  {NoStop}%
\bibitem [{\citenamefont {Salvat}\ \emph {et~al.}(2005)\citenamefont {Salvat},
  \citenamefont {Jablonski},\ and\ \citenamefont {Powell}}]{SalvatELSEPA2005}%
  \BibitemOpen
  \bibfield  {author} {\bibinfo {author} {\bibfnamefont {F.}~\bibnamefont
  {Salvat}}, \bibinfo {author} {\bibfnamefont {A.}~\bibnamefont {Jablonski}}, \
  and\ \bibinfo {author} {\bibfnamefont {C.~J.}\ \bibnamefont {Powell}},\
  }\href {\doibase 10.1016/j.cpc.2004.09.006} {\bibfield  {journal} {\bibinfo
  {journal} {Comp. Phys. Comm.}\ }\textbf {\bibinfo {volume} {165}},\ \bibinfo
  {pages} {157 } (\bibinfo {year} {2005})}\BibitemShut {NoStop}%
\bibitem [{\citenamefont {Salvat}\ and\ \citenamefont
  {Fern\'andez-Varea}(2019)}]{SalvatRadial2019}%
  \BibitemOpen
  \bibfield  {author} {\bibinfo {author} {\bibfnamefont {F.}~\bibnamefont
  {Salvat}}\ and\ \bibinfo {author} {\bibfnamefont {J.~M.}\ \bibnamefont
  {Fern\'andez-Varea}},\ }\href {\doibase
  https://doi.org/10.1016/j.cpc.2019.02.011} {\bibfield  {journal} {\bibinfo
  {journal} {Computer Physics Communications}\ }\textbf {\bibinfo {volume}
  {240}},\ \bibinfo {pages} {165} (\bibinfo {year} {2019})}\BibitemShut
  {NoStop}%
\bibitem [{\citenamefont {Yennie}\ \emph {et~al.}(1954)\citenamefont {Yennie},
  \citenamefont {Ravenhall},\ and\ \citenamefont
  {Wilson}}]{Yennie.PhaseShiftSum.1954}%
  \BibitemOpen
  \bibfield  {author} {\bibinfo {author} {\bibfnamefont {D.~R.}\ \bibnamefont
  {Yennie}}, \bibinfo {author} {\bibfnamefont {D.~G.}\ \bibnamefont
  {Ravenhall}}, \ and\ \bibinfo {author} {\bibfnamefont {R.~N.}\ \bibnamefont
  {Wilson}},\ }\href {\doibase 10.1103/PhysRev.95.500} {\bibfield  {journal}
  {\bibinfo  {journal} {Phys. Rev.}\ }\textbf {\bibinfo {volume} {95}},\
  \bibinfo {pages} {500} (\bibinfo {year} {1954})}\BibitemShut {NoStop}%
\bibitem [{\citenamefont {Tarrach}(1975)}]{Tarrach1975}%
  \BibitemOpen
  \bibfield  {author} {\bibinfo {author} {\bibfnamefont {R.}~\bibnamefont
  {Tarrach}},\ }\href {\doibase 10.1007/BF02894857} {\bibfield  {journal}
  {\bibinfo  {journal} {Nuovo Cimento Soc. Ital. Fis.}\ }\textbf {\bibinfo
  {volume} {28}},\ \bibinfo {pages} {409} (\bibinfo {year} {1975})}\BibitemShut
  {NoStop}%
\bibitem [{\citenamefont {Drechsel}\ \emph {et~al.}(1997)\citenamefont
  {Drechsel}, \citenamefont {Kn\"ochlein}, \citenamefont {Metz},\ and\
  \citenamefont {Scherer}}]{DrechselVVCS-spin0-1997}%
  \BibitemOpen
  \bibfield  {author} {\bibinfo {author} {\bibfnamefont {D.}~\bibnamefont
  {Drechsel}}, \bibinfo {author} {\bibfnamefont {G.}~\bibnamefont
  {Kn\"ochlein}}, \bibinfo {author} {\bibfnamefont {A.}~\bibnamefont {Metz}}, \
  and\ \bibinfo {author} {\bibfnamefont {S.}~\bibnamefont {Scherer}},\ }\href
  {\doibase 10.1103/PhysRevC.55.424} {\bibfield  {journal} {\bibinfo  {journal}
  {Phys. Rev. C}\ }\textbf {\bibinfo {volume} {55}},\ \bibinfo {pages} {424}
  (\bibinfo {year} {1997})}\BibitemShut {NoStop}%
\bibitem [{\citenamefont {Lensky}\ \emph {et~al.}(2018)\citenamefont {Lensky},
  \citenamefont {Hagelstein}, \citenamefont {Pascalutsa},\ and\ \citenamefont
  {Vanderhaeghen}}]{Lensky-VVCS-Spin0-2018}%
  \BibitemOpen
  \bibfield  {author} {\bibinfo {author} {\bibfnamefont {V.}~\bibnamefont
  {Lensky}}, \bibinfo {author} {\bibfnamefont {F.}~\bibnamefont {Hagelstein}},
  \bibinfo {author} {\bibfnamefont {V.}~\bibnamefont {Pascalutsa}}, \ and\
  \bibinfo {author} {\bibfnamefont {M.}~\bibnamefont {Vanderhaeghen}},\ }\href
  {\doibase 10.1103/PhysRevD.97.074012} {\bibfield  {journal} {\bibinfo
  {journal} {Phys. Rev. D}\ }\textbf {\bibinfo {volume} {97}},\ \bibinfo
  {pages} {074012} (\bibinfo {year} {2018})}\BibitemShut {NoStop}%
\bibitem [{\citenamefont {Aleksanian}\ \emph {et~al.}(1987)\citenamefont
  {Aleksanian} \emph {et~al.}}]{Aleksanian:1986hb}%
  \BibitemOpen
  \bibfield  {author} {\bibinfo {author} {\bibfnamefont {A.}~\bibnamefont
  {Aleksanian}} \emph {et~al.},\ }\href@noop {} {\bibfield  {journal} {\bibinfo
   {journal} {Sov. J. Nucl. Phys.}\ }\textbf {\bibinfo {volume} {45}},\
  \bibinfo {pages} {628} (\bibinfo {year} {1987})}\BibitemShut {NoStop}%
\bibitem [{\citenamefont {Criegee}\ \emph {et~al.}(1977)\citenamefont
  {Criegee}, \citenamefont {Franke}, \citenamefont {Giese}, \citenamefont
  {Khal}, \citenamefont {Poelz}, \citenamefont {Timm}, \citenamefont {Werner},\
  and\ \citenamefont {Zimmermann}}]{Criegee:1977uf}%
  \BibitemOpen
  \bibfield  {author} {\bibinfo {author} {\bibfnamefont {L.}~\bibnamefont
  {Criegee}}, \bibinfo {author} {\bibfnamefont {G.}~\bibnamefont {Franke}},
  \bibinfo {author} {\bibfnamefont {A.}~\bibnamefont {Giese}}, \bibinfo
  {author} {\bibfnamefont {T.}~\bibnamefont {Khal}}, \bibinfo {author}
  {\bibfnamefont {G.}~\bibnamefont {Poelz}}, \bibinfo {author} {\bibfnamefont
  {U.}~\bibnamefont {Timm}}, \bibinfo {author} {\bibfnamefont {H.}~\bibnamefont
  {Werner}}, \ and\ \bibinfo {author} {\bibfnamefont {W.}~\bibnamefont
  {Zimmermann}},\ }\href {\doibase 10.1016/0550-3213(77)90325-X} {\bibfield
  {journal} {\bibinfo  {journal} {Nucl. Phys. B}\ }\textbf {\bibinfo {volume}
  {121}},\ \bibinfo {pages} {38} (\bibinfo {year} {1977})}\BibitemShut
  {NoStop}%
\bibitem [{\citenamefont {Roca-Maza}(2017)}]{Roca_Maza_2017}%
  \BibitemOpen
  \bibfield  {author} {\bibinfo {author} {\bibfnamefont {X.}~\bibnamefont
  {Roca-Maza}},\ }\href {\doibase 10.1209/0295-5075/120/33002} {\bibfield
  {journal} {\bibinfo  {journal} {{EPL} (Europhysics Letters)}\ }\textbf
  {\bibinfo {volume} {120}},\ \bibinfo {pages} {33002} (\bibinfo {year}
  {2017})}\BibitemShut {NoStop}%
\bibitem [{\citenamefont {{De Jager}}\ \emph {et~al.}(1974)\citenamefont {{De
  Jager}}, \citenamefont {{De Vries}},\ and\ \citenamefont {{De
  Vries}}}]{Vries1974}%
  \BibitemOpen
  \bibfield  {author} {\bibinfo {author} {\bibfnamefont {C.}~\bibnamefont {{De
  Jager}}}, \bibinfo {author} {\bibfnamefont {H.}~\bibnamefont {{De Vries}}}, \
  and\ \bibinfo {author} {\bibfnamefont {C.}~\bibnamefont {{De Vries}}},\
  }\href {\doibase https://doi.org/10.1016/S0092-640X(74)80002-1} {\bibfield
  {journal} {\bibinfo  {journal} {Atomic Data and Nuclear Data Tables}\
  }\textbf {\bibinfo {volume} {14}},\ \bibinfo {pages} {479 } (\bibinfo {year}
  {1974})},\ \bibinfo {note} {nuclear Charge and Moment
  Distributions}\BibitemShut {NoStop}%
\bibitem [{\citenamefont {Gorchtein}\ \emph {et~al.}(2011)\citenamefont
  {Gorchtein}, \citenamefont {Hobbs}, \citenamefont {Londergan},\ and\
  \citenamefont {Szczepaniak}}]{Gorchtein:2011xx}%
  \BibitemOpen
  \bibfield  {author} {\bibinfo {author} {\bibfnamefont {M.}~\bibnamefont
  {Gorchtein}}, \bibinfo {author} {\bibfnamefont {T.}~\bibnamefont {Hobbs}},
  \bibinfo {author} {\bibfnamefont {J.}~\bibnamefont {Londergan}}, \ and\
  \bibinfo {author} {\bibfnamefont {A.~P.}\ \bibnamefont {Szczepaniak}},\
  }\href {\doibase 10.1103/PhysRevC.84.065202} {\bibfield  {journal} {\bibinfo
  {journal} {Phys. Rev. C}\ }\textbf {\bibinfo {volume} {84}},\ \bibinfo
  {pages} {065202} (\bibinfo {year} {2011})},\ \Eprint
  {http://arxiv.org/abs/1110.5982} {arXiv:1110.5982 [nucl-th]} \BibitemShut
  {NoStop}%
\bibitem [{\citenamefont {Caldwell}\ \emph {et~al.}(1973)\citenamefont
  {Caldwell}, \citenamefont {Elings}, \citenamefont {Hesse}, \citenamefont
  {Morrison}, \citenamefont {Murphy},\ and\ \citenamefont
  {Yount}}]{Caldwell:1973bu}%
  \BibitemOpen
  \bibfield  {author} {\bibinfo {author} {\bibfnamefont {D.~O.}\ \bibnamefont
  {Caldwell}}, \bibinfo {author} {\bibfnamefont {V.}~\bibnamefont {Elings}},
  \bibinfo {author} {\bibfnamefont {W.}~\bibnamefont {Hesse}}, \bibinfo
  {author} {\bibfnamefont {R.}~\bibnamefont {Morrison}}, \bibinfo {author}
  {\bibfnamefont {F.~V.}\ \bibnamefont {Murphy}}, \ and\ \bibinfo {author}
  {\bibfnamefont {D.}~\bibnamefont {Yount}},\ }\href {\doibase
  10.1103/PhysRevD.7.1362} {\bibfield  {journal} {\bibinfo  {journal} {Phys.
  Rev. D}\ }\textbf {\bibinfo {volume} {7}},\ \bibinfo {pages} {1362} (\bibinfo
  {year} {1973})}\BibitemShut {NoStop}%
\bibitem [{\citenamefont {Caldwell}\ \emph {et~al.}(1979)\citenamefont
  {Caldwell} \emph {et~al.}}]{Caldwell:1978ik}%
  \BibitemOpen
  \bibfield  {author} {\bibinfo {author} {\bibfnamefont {D.~O.}\ \bibnamefont
  {Caldwell}} \emph {et~al.},\ }\href {\doibase 10.1103/PhysRevLett.42.553}
  {\bibfield  {journal} {\bibinfo  {journal} {Phys. Rev. Lett.}\ }\textbf
  {\bibinfo {volume} {42}},\ \bibinfo {pages} {553} (\bibinfo {year}
  {1979})}\BibitemShut {NoStop}%
\bibitem [{\citenamefont {Bianchi}\ \emph {et~al.}(1996)\citenamefont
  {Bianchi}, \citenamefont {Muccifora}, \citenamefont {De~Sanctis},
  \citenamefont {Fantoni}, \citenamefont {Levi~Sandri}, \citenamefont {Polli},
  \citenamefont {Reolon}, \citenamefont {Rossi}, \citenamefont {Anghinolfi},
  \citenamefont {Corvisiero}, \citenamefont {Ripani}, \citenamefont {Sanzone},
  \citenamefont {Taiuti},\ and\ \citenamefont
  {Zucchiatti}}]{Bianchi-Photoabsorption-1996}%
  \BibitemOpen
  \bibfield  {author} {\bibinfo {author} {\bibfnamefont {N.}~\bibnamefont
  {Bianchi}}, \bibinfo {author} {\bibfnamefont {V.}~\bibnamefont {Muccifora}},
  \bibinfo {author} {\bibfnamefont {E.}~\bibnamefont {De~Sanctis}}, \bibinfo
  {author} {\bibfnamefont {A.}~\bibnamefont {Fantoni}}, \bibinfo {author}
  {\bibfnamefont {P.}~\bibnamefont {Levi~Sandri}}, \bibinfo {author}
  {\bibfnamefont {E.}~\bibnamefont {Polli}}, \bibinfo {author} {\bibfnamefont
  {A.~R.}\ \bibnamefont {Reolon}}, \bibinfo {author} {\bibfnamefont
  {P.}~\bibnamefont {Rossi}}, \bibinfo {author} {\bibfnamefont
  {M.}~\bibnamefont {Anghinolfi}}, \bibinfo {author} {\bibfnamefont
  {P.}~\bibnamefont {Corvisiero}}, \bibinfo {author} {\bibfnamefont
  {M.}~\bibnamefont {Ripani}}, \bibinfo {author} {\bibfnamefont
  {M.}~\bibnamefont {Sanzone}}, \bibinfo {author} {\bibfnamefont
  {M.}~\bibnamefont {Taiuti}}, \ and\ \bibinfo {author} {\bibfnamefont
  {A.}~\bibnamefont {Zucchiatti}},\ }\href {\doibase 10.1103/PhysRevC.54.1688}
  {\bibfield  {journal} {\bibinfo  {journal} {Phys. Rev. C}\ }\textbf {\bibinfo
  {volume} {54}},\ \bibinfo {pages} {1688} (\bibinfo {year}
  {1996})}\BibitemShut {NoStop}%
\bibitem [{\citenamefont {Christy}\ and\ \citenamefont
  {Bosted}(2010)}]{ChristyFit2010}%
  \BibitemOpen
  \bibfield  {author} {\bibinfo {author} {\bibfnamefont {M.~E.}\ \bibnamefont
  {Christy}}\ and\ \bibinfo {author} {\bibfnamefont {P.~E.}\ \bibnamefont
  {Bosted}},\ }\href {\doibase 10.1103/PhysRevC.81.055213} {\bibfield
  {journal} {\bibinfo  {journal} {Phys. Rev. C}\ }\textbf {\bibinfo {volume}
  {81}},\ \bibinfo {pages} {055213} (\bibinfo {year} {2010})}\BibitemShut
  {NoStop}%
\bibitem [{\citenamefont {Souder}\ \emph {et~al.}(2011)\citenamefont {Souder},
  \citenamefont {Holmes}, \citenamefont {Jen}, \citenamefont {Zana},
  \citenamefont {Ahmed}, \citenamefont {Rakhman}, \citenamefont {Cisbani},
  \citenamefont {Frullani}, \citenamefont {Garibaldi}, \citenamefont {Meddi}
  \emph {et~al.}}]{PREX-II}%
  \BibitemOpen
  \bibfield  {author} {\bibinfo {author} {\bibfnamefont {P.}~\bibnamefont
  {Souder}}, \bibinfo {author} {\bibfnamefont {R.}~\bibnamefont {Holmes}},
  \bibinfo {author} {\bibfnamefont {C.-M.}\ \bibnamefont {Jen}}, \bibinfo
  {author} {\bibfnamefont {L.}~\bibnamefont {Zana}}, \bibinfo {author}
  {\bibfnamefont {Z.}~\bibnamefont {Ahmed}}, \bibinfo {author} {\bibfnamefont
  {A.}~\bibnamefont {Rakhman}}, \bibinfo {author} {\bibfnamefont
  {E.}~\bibnamefont {Cisbani}}, \bibinfo {author} {\bibfnamefont
  {S.}~\bibnamefont {Frullani}}, \bibinfo {author} {\bibfnamefont
  {F.}~\bibnamefont {Garibaldi}}, \bibinfo {author} {\bibfnamefont
  {F.}~\bibnamefont {Meddi}},  \emph {et~al.},\ }\href
  {http://hallaweb.jlab.org/parity/prex/prexII.pdf} {\bibfield  {journal}
  {\bibinfo  {journal} {Proposal for Jefferson Lab PAC}\ }\textbf {\bibinfo
  {volume} {38}} (\bibinfo {year} {2011})}\BibitemShut {NoStop}%
\bibitem [{\citenamefont {Mammei}\ \emph {et~al.}(2013)\citenamefont {Mammei},
  \citenamefont {McNulty}, \citenamefont {Michaels}, \citenamefont {Paschke},
  \citenamefont {Riordan}, \citenamefont {Souder} \emph {et~al.}}]{CREX}%
  \BibitemOpen
  \bibfield  {author} {\bibinfo {author} {\bibfnamefont {J.}~\bibnamefont
  {Mammei}}, \bibinfo {author} {\bibfnamefont {D.}~\bibnamefont {McNulty}},
  \bibinfo {author} {\bibfnamefont {R.}~\bibnamefont {Michaels}}, \bibinfo
  {author} {\bibfnamefont {K.}~\bibnamefont {Paschke}}, \bibinfo {author}
  {\bibfnamefont {S.}~\bibnamefont {Riordan}}, \bibinfo {author} {\bibfnamefont
  {P.}~\bibnamefont {Souder}},  \emph {et~al.},\ }\href
  {http://hallaweb.jlab.org/parity/prex/c-rex2013_v7.pdf} {\bibfield  {journal}
  {\bibinfo  {journal} {Proposal to Jefferson Lab PAC}\ }\textbf {\bibinfo
  {volume} {39}} (\bibinfo {year} {2013})}\BibitemShut {NoStop}%
\bibitem [{\citenamefont {Berman}\ and\ \citenamefont
  {Fultz}(1975)}]{Berman:1975tt}%
  \BibitemOpen
  \bibfield  {author} {\bibinfo {author} {\bibfnamefont {B.}~\bibnamefont
  {Berman}}\ and\ \bibinfo {author} {\bibfnamefont {S.}~\bibnamefont {Fultz}},\
  }\href {\doibase 10.1103/RevModPhys.47.713} {\bibfield  {journal} {\bibinfo
  {journal} {Rev. Mod. Phys.}\ }\textbf {\bibinfo {volume} {47}},\ \bibinfo
  {pages} {713} (\bibinfo {year} {1975})}\BibitemShut {NoStop}%
\bibitem [{\citenamefont {Adhikari}\ \emph {et~al.}(2021)\citenamefont
  {Adhikari} \emph {et~al.}}]{Adhikari:2021phr}%
  \BibitemOpen
  \bibfield  {author} {\bibinfo {author} {\bibfnamefont {D.}~\bibnamefont
  {Adhikari}} \emph {et~al.},\ }\href@noop {} {\  (\bibinfo {year} {2021})},\
  \Eprint {http://arxiv.org/abs/2102.10767} {arXiv:2102.10767 [nucl-ex]}
  \BibitemShut {NoStop}%
\bibitem [{\citenamefont {Kumar}(2020)}]{Kumar:2020ejz}%
  \BibitemOpen
  \bibfield  {author} {\bibinfo {author} {\bibfnamefont {K.~S.}\ \bibnamefont
  {Kumar}} (\bibinfo {collaboration} {PREX, CREX}),\ }\href {\doibase
  10.1016/j.aop.2019.168012} {\bibfield  {journal} {\bibinfo  {journal} {Annals
  Phys.}\ }\textbf {\bibinfo {volume} {412}},\ \bibinfo {pages} {168012}
  (\bibinfo {year} {2020})}\BibitemShut {NoStop}%
\bibitem [{\citenamefont {Armstrong}()}]{Qweak_Bn}%
  \BibitemOpen
  \bibfield  {author} {\bibinfo {author} {\bibfnamefont {D.}~\bibnamefont
  {Armstrong}},\ }\href@noop {} {}\bibinfo {howpublished} {private
  communication}\BibitemShut {NoStop}%
\end{thebibliography}
%

\end{document}